\pdfoutput=1
\documentclass[12pt]{iopart}

\usepackage{color}
\definecolor{indigo}{RGB}{0,0,120}
\usepackage[colorlinks=true, linkcolor=indigo, citecolor=blue, urlcolor=indigo]{hyperref}

\usepackage{graphicx}%
\usepackage{dcolumn}%
\usepackage{setspace}

\usepackage{iopams}  
\begin{document}

\title[Simulation of the internal kink mode in visco-resistive regimes]{Simulation of the internal kink mode in visco-resistive regimes}%
\author{J. Mendonca}
 \address{Institute for Plasma Research, Bhat, Gandhinagar 382428, India}%
 \address{Homi Bhabha National Institute, Training School Complex, Anushakti Nagar, Mumbai -400094, India}
\author{D. Chandra}%
\address{Institute for Plasma Research, Bhat, Gandhinagar 382428, India%
}%
\address{Homi Bhabha National Institute, Training School Complex, Anushakti Nagar, Mumbai -400094, India}

\author{A. Sen}
\address{%
Institute for Plasma Research, Bhat, Gandhinagar 382428, India%
}%
\address{Homi Bhabha National Institute, Training School Complex, Anushakti Nagar, Mumbai -400094, India}

\author{A.Thyagaraja}
 \address{Astrophysics Group, University of Bristol, Bristol, BS8 1TL, UK}

\ead{jervis.mendonca@ipr.res.in}
\vspace{10pt}

\begin{abstract}
We present numerical simulation results of the nonlinear evolution of the (1,1) internal kink mode in the presence of various kinds of equilibrium plasma flows. The present studies are carried out in the framework of a two fluid model to extend our past investigations done with a reduced magneto-hydrodynamic (RMHD) model. 
Two-fluid effects are found to significantly influence the mode dynamics in a number of different ways. In the linear regime diamagnetic effects in combination with flows provide a synergistic stabilizing influence that also carries over to the nonlinear regime. In addition one observes novel symmetry breaking phenomena in the linear growth rates as well as in the nonlinear saturated states of the mode. Our study also explores the influence of strong viscosity on the mode evolution and finds interesting modifications in the real frequency of the mode in the linear regime.
\end{abstract}

%

%
%
%
%

\section{Introduction}\label{introduction}
In an advanced tokamak with high temperature plasmas, sawtooth is a frequently observed phenomenon. As the core temperature increases, resistivity goes down which results in a higher current density in the core region. Simultaneously, the safety factor, $q$, becomes less than one which triggers an $m=1$, $n=1$ internal kink mode. As sawteeth can potentially degrade the confinement of the plasma, it is very important to understand its dynamics and learn how to control it. To do that, we first need to understand the $m=1$, $n=1$ mode, as it is closely associated with sawteeth. In view of its fundamental importance the mode has been widely studied in the past to understand its physics \cite{Ara1978,Thyagaraja1991,Bondeson1992}. Plasma rotation is also a very common occurrence in a tokamak that can get generated either internally or  externally e.g. due to NBI injection. It is also well known that plasma rotation can modify sawtooth dynamics\cite{Chandra2005,Thyagaraja2010,Chapman2010,Chandra2015,Mendonca2018}. Experimental observations in NSTX \cite{Menard2003,Menard2005}, JET  \cite{Chapman2007a, Nave2006}, MAST \cite{Chapman2006a}, TEXTOR\cite{Chapman2008c} etc. indicate that there is a definite correlation between the changes in sawtooth behaviour with changes of plasma rotation in a tokamak. In particular it is seen that the direction of the flow can either increase or decrease the sawtooth period. In other words flow can change the intrinsic characteristics of the kink mode. There are several past studies that have focused on the stability of the $m=1$,$n=1$ mode in the presence of flow. 
Guzdar et al\cite{Guzdar1234} have shown that toroidal sheared flow which is close to the acoustic speed in plasma can completely stabilise the (1,1) mode. In the work of  Shumlak et al.\cite{Shumlak1995}, a stabilising effect due to a sheared axial flow on the (1,1) mode in a cylindrical Z-pinch has been found. Most of the past studies found stabilising effects of flow on the (1,1) mode, but there are a few studies such as by Gatto et al\cite{gatto} and Brunetti et al \cite{Brunetti2017}, Crombe et al\cite{crombe} that have found that flow shear can have a destabilising effect. Chapman et al\cite{Chapman2006a} have suggested that the asymmetry in the sawteeth period is related to the plasma flow direction with respect to the diamagnetic drift. In our earlier study\cite{Mendonca2018}, using a single fluid model, we have observed that there is a similar symmetry breaking in the presence of helical flows. However, those single fluid studies do not include diamagnetic flows which require a two fluid model.  Diamagnetic flows can alter the dynamics of the mode, particularly in the presence of equilibrium flows, by affecting  its stability as well as its rotation frequency.  \\ 
A number of past studies have employed the two fluid model to study of tokamak instabilities, particularly the tearing and kink instabilities \cite{Barkov,zakharov,peteryoon,daughton,cihan}. One of the earliest works on this topic has been by Thyagaraja et al\cite{Thyagaraja1991}, in which they have elucidated the necessary and sufficient conditions required for the existence of a nonlinearly saturated m = 1 tearing mode in tokamaks. In the paper of Barkov et al\cite{Barkov}, they have studied the drift-kink instabilities using two fluid simulations and compared their accuracy to that of PIC methods and found good agreement. Zakharov et al\cite{zakharov} have studied the internal kink mode in the context of tokamaks. They studied a scenario with a finite $\beta$ and found their results to compare favorably with past kinetic studies. A validation study using the NIMROD code was carried out by Akcay et al\cite{cihan}. Although we have several interesting results on the effect of flows on the dynamics of the $m=1$,$n=1$ internal kink modes with flows using a single fluid model, those results can get modified in the two fluid regime. Our present study is aimed at studying those modifications  by  extending our earlier single fluid study\cite{Mendonca2018} to a two fluid regime and thereby gaining a deeper understanding of the dynamics of the internal kink mode.

In this work, we have investigated the stability of the (1,1) mode in the presence of sheared flows over a range of viscosity regimes using the CUTIE\cite{Thyagaraja2000} code in the two fluid regime. The reason for taking a
wide range of viscosity is to account for the fact that viscosoity can sometimes be high in tokamak operations, possibly due  to turbulent effects and could therefore modify the effect of flow shear on the stability of the internal kink mode \cite{itoh,Takeda2008,Mendonca2018} as observed earlier. We also take account of the electron  diamagnetic drift velocity, $v_{d}$ and the corresponding drift frequency $\omega*$. The diamagnetic drift is proportional to the electron density gradient which is characterized  by the parameter $\alpha$ in our model where we have used a density profile of the form $n=n_{0}exp({-\alpha\frac{r^2}{a^2}})$, where $n$ stands for density and $r$ is the radial coordinate.

We have begun with linear studies which we have carried out using the Resolvent method, a method of finding eigenvalues, explained in the paper\cite{Chandra2015}. We have observed the modification of growth rate and the rotation frequency of the $m=1$,$n=1$ with different diamagnetic flow frequencies over a range of viscosities. Then we have applied pure axial, pure poloidal and helical flows, and studied the same for different combinations of flow magnitudes and directions. In all cases we have seen symmetry breaking with a reversal of flow direction. This symmetry breaking happens either in the presence of diamagnetic flow or by including parallel dynamics or both. However, this dynamics is significantly modified depending upon the viscosity regime. Then we have extended our work into the nonlinear regime and observed the nonlinear saturation level to be similarly modified with diamagnetic flows, imposed flows, as well as viscosity. Both the linear and nonlinear results are very different as compared to our studies in the RMHD regime \cite{Mendonca2018}.  We also notice that the poloidal flow is destabilising nonlinearly and linearly in some cases.\\

The paper is organized in the following manner. In section \ref{model}, we have described the two fluid model of plasma in a cylindrical geometry. In section \ref{linear results}, we have studied the effect of diamagnetic flows  (1,1) mode in the linear regime without any imposed flows initially. After that, we have studied the behaviour of the mode with different types of flows. We have done these studies both in the low and high viscosity regimes. In section \ref{nonlinear results} we have  studied the (1,1) mode  in the nonlinear regime in the absence of flow as well as in presence of axial, poloidal and helical flows. Section \ref{discussion} provides a brief summary and a discussion of the results.

\section{Model} \label{model}

We have previously published our investigations using a
single-fluid incompressible version of the CUTIE code \cite{Mendonca2018}. In the present version, our numerical investigations have been carried out in the framework of a two fluid model, that includes a continuity equation for electron density and the parallel momentum equation. We use a periodic cylinder geometry $(\rho,\theta,z)$,($\rho$ being the radial coordinate, $\theta$ being the azimuthal coordinate, and $z$ being the axial coordinate) defined in terms of the minor radius, $a$, and  the major radius, $R_{0}$ as follows: we set $\rho = r/a$, $r$ being the radial distance, $0\leqslant\rho\leqslant1;0\leqslant\theta,\zeta\leqslant2\pi$  ;$\zeta=z/R_{0}$, is analogous to the toroidal angle. This model thus includes drift effects of two fluid theory, containing density and parallel momentum effects in addition to the single fluid model. We utilise CGS electrostatic units. We have neglected curvature effects as we are using a large aspect ratio, and as in our previous work, toroidal coupling between different $(m,n)$ modes is allowed. We use a constant, uniform axial magnetic field, while the poloidal component of the magnetic field is determined as we evolve the equations. We prescribe and hold fixed imposed equilibrium flows and density profiles. Also, plasma $\beta$ is assumed to be low and other code parameters are chosen to be consistent with this assumption. A significant difference is that we allow the current profile to evolve in the nonlinear evolution. The equations in our model are as follows\cite{Thyagaraja2004,Thyagaraja2010}:

\begin{equation}\label{mhd1}
\frac{\partial n}{\partial t}+\nabla\cdot\left(n\mathbf{v}\right)=S_{p} 
\end{equation}

\begin{equation}\label{mhd2}
m_{i}n\frac{d\mathbf{v}}{d\mathit{t}}=-\nabla p+\frac{\mathbf{j\times B}}{c}+\mathbf{F}_{eff}
\end{equation}

\begin{equation}\label{mhd3}
\frac{3}{2}\frac{dp_{i,e}}{d\mathit{t}}+p_{i,e}\nabla\cdot\mathbf{v_{i,e}}=-\nabla\cdot\mathbf{q_{i,e}}+P_{i,e}
\end{equation}

\begin{equation}\label{mhd4}
\mathbf{E}+\frac{\mathbf{v_{e}\times B}}{c}=-\frac{\nabla p_{e}}{en}+\mathbf{R_{e}}
\end{equation}

\begin{equation}\label{mhd5}
\nabla\times\mathbf{B}=\frac{4\pi\mathbf{j}}{c}
\end{equation}

\begin{equation}\label{mhd6}
\nabla \times \mathbf{E}=-\frac{1}{c}\frac{\partial \mathbf{B}}{\partial t} 
\end{equation}

It must be mentioned that the form of the source terms has been described in earlier papers\cite{Thyagaraja2004,Thyagaraja2010}. We use a reduced model for our present purposes.  We note here that $T_{e}=T_{i}=T_{0}$. We will introduce the dependent variables here, namely $\phi$, the electrostatic potential, $\psi$, poloidal flux function, $n_{e}$, quasi-neutral electron number density, $v_{\parallel}$, parallel velocity. These are functions of $r$, $\theta$, $z$, $t$, or in the normalized form $\rho$, $\theta$, $\zeta$ and $t$ as introduced above. We can write these by splitting the equilibrium and fluctuation parts, or in other words by Fourier analysing them and keeping the (0,0) component separate. Thus,

\begin{equation}
F\left(\rho,\theta,\zeta,t\right)=F_{0}\left(\rho,t\right)+\tilde{f}\left(\rho,\theta,\zeta,t\right)
\end{equation}
 
Here, 
\begin{equation}
F_{0}\left(\rho,t\right)=\intop_{0}^{2\pi}\frac{d\theta}{2\pi}\intop_{0}^{2\pi}\frac{d\zeta}{2\pi}F\left(\rho,\theta,\zeta,t\right)
\end{equation} and 

\begin{equation}
\tilde{f}\left(\rho,\theta,\zeta,t\right)=\sum_{m}\sum_{n}\hat{f}_{m,n}\left(\rho,t\right)e^{i\left(m\theta+n\zeta\right)}
\end{equation}

We also introduce the fundamental equilibrium quantities of the model: the axial magnetic field, $B_{tor}=B_{\zeta}=B_{0}$(uniform and constant),the safety factor, $q(\rho)=\rho\left(\frac{a}{R_{0}}\right)\frac{B_{0}}{B_{0\theta}\left(\rho\right)}$, equilibrium quasi-neutral electron density $n_{0}\left(\rho\right)$. They have the following profiles:

\begin{equation}
q\left(\rho\right)=q_{0}\left[1+\left(\frac{\rho}{\rho_{0}}\right)^{2\Lambda}\right]^{1/\Lambda}
\end{equation}

\begin{figure}[h]
\includegraphics[scale=0.4]{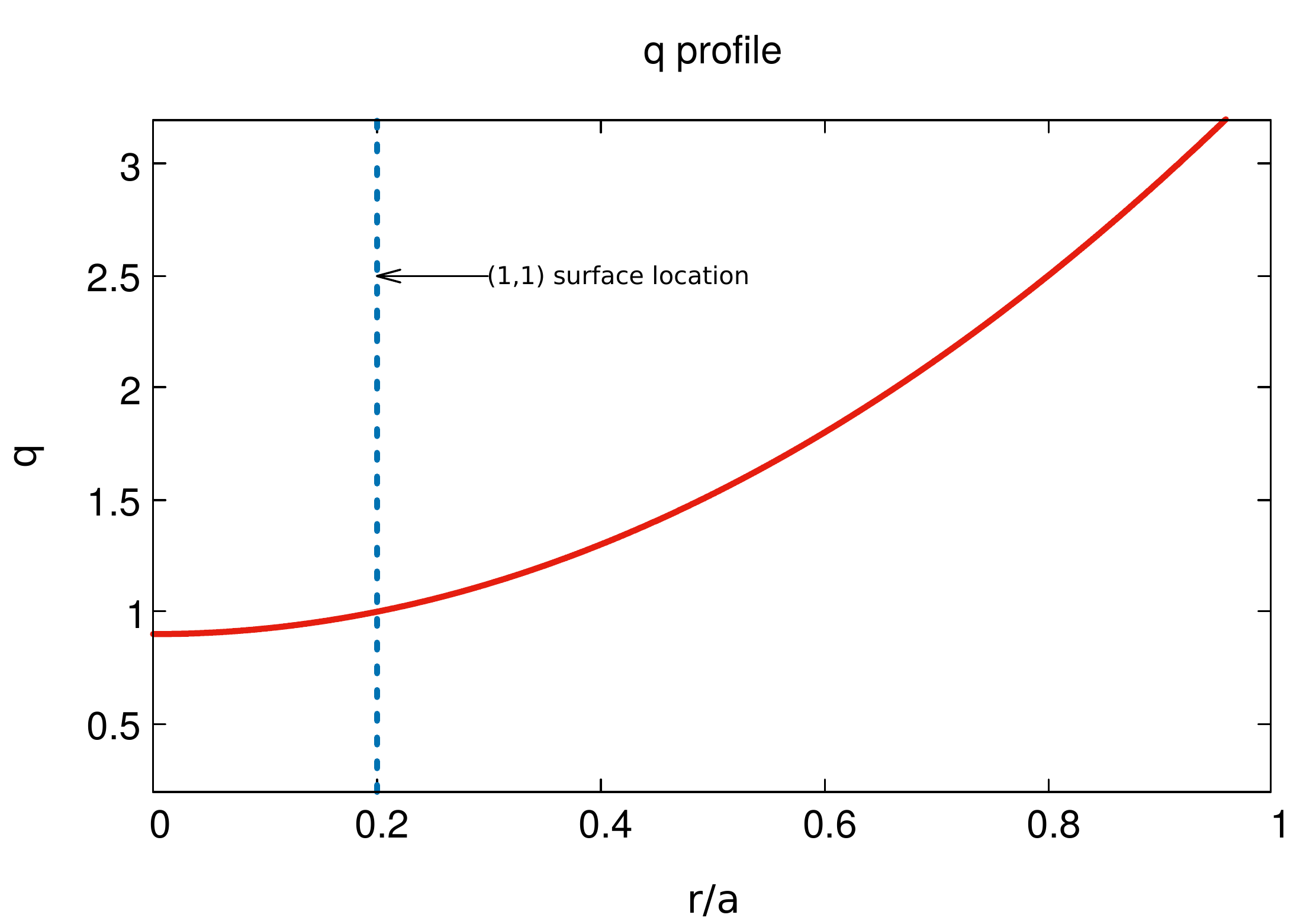}    
\caption{q profile}
\label{fig:q_profile}
\end{figure}

Fig. \ref{fig:q_profile} shows the $q$ profile used in the simulations and the $q=1$ surface.

\begin{equation}
n_{0}\left(\rho\right)=n_{0}\left(0\right)e^{-\alpha\rho^{2}}
\end{equation}

The values used for these parameters are as follows, consistent with our earlier single fluid study\cite{Mendonca2018}:
$q_{0}=0.9$, $\Lambda=1$, $\rho_{0}=0.6$, $T_{0}=275$eV,       $B_{0}=2\times10^4$,    $I_{p}=29.4KA$.
The parameter $\alpha$ is proportional to the density gradient, and we vary it from $\alpha=0$, which is the single fluid case, to $\alpha=1.5$.
 
We will now write the equations for the fluctuating quantities. We keep in mind that the following variables are to be evolved:

\begin{itemize}
\item The fluctuating quantities, $\tilde{n}$, $\tilde{\phi}$,$\tilde{\psi}$, $\tilde{v}_{\parallel}$, $\tilde{W}$. Here, $\tilde{W}$ is the linearised `potential vorticity'. The mean flows are held fixed, namely $v_{z0}$ and $v_{\theta0}$. 
\item The mean poloidal field $B_{p}(r,t)$ in the nonlinear case.
\end{itemize}

It is convenient to define the following time-scales which are relevant to our work: $\tau_{A}=a/v_{A}$ is the Alfv\'{e}n time; $\tau_{\eta}=(4\pi a^{2}/c^{2}\eta)$ the resistive diffusion time; $\tau_{\nu}=(a^{2}/\nu)$ the viscous diffusion time. We will use in the following the \textit{Lundquist Number}, $S=\frac{\tau_{\eta}}{\tau_{A}}$ and the \textit{Prandtl Number}, $Pr=\frac{\tau_{\eta}}{\tau_{\nu}}$. The velocity perturbations are non-dimensionalised relative to the Alfven speed, $v_{A}=\frac{B_{0}}{\left(4\pi m_{i}n_{0}\right)^{1/2}}$, and thermal velocity, $V_{TH}=\left(\frac{T_{e}\left(0,t\right)+T_{i}\left(0,t\right)}{m_{i}}\right)^{1/2}$. Additionally, $\rho_{s}=\frac{v_{th}}{\omega_{ci}}$,  where,  $v_{th}^{2}=\left(T_{0i}+T_{0e}\right)/m_{i}$,  $\omega_{ci}=\left(eB_{0}/m_{i}c\right)$, with $T_{0i},T_{0e}$ being ion and electron temperatures respectively. $m_{i}$ is the ion mass, $e$ is the elementary charge. $\Phi_{0}(r), \Psi_{0}(r)$ denote the mean electrostatic and magnetostatic potentials respectively. The equilibrium axial and poloidal, sub-Alfv\'{e}nic sheared flows are: $M_{z}=V_{0z}/v_{A}$ is the Axial Mach number; $M_{\theta}(\rho)=\rho\frac{a\Omega(\rho)}{v_{A}}$ is the poloidal Mach number.

These equations are rewritten in terms of the variables:

\begin{eqnarray}
\label{Eq.1}
\fl \tilde{W}=\rho_{s}^{2}\nabla\cdot\left(\frac{n_{0}\left(\rho\right)}{n_{0}\left(0\right)}\nabla_{\perp}\tilde{\phi}\right)
\end{eqnarray}

\begin{eqnarray}
\label{Eq.2}
\fl \frac{\partial\tilde{W}}{\partial t}+\mathrm{\mathbf{v_{0}}}\cdot\nabla\tilde{W}+  v_{A}\nabla_{\parallel}\rho_{s}^{2}\nabla_{\perp}^{2}\tilde{\psi} = \nonumber\\
v_{A}\rho_{s}\frac{1}{r}\frac{\partial\tilde{\psi}}{\partial\theta}\frac{4\pi\rho_{s}}{cB_{0}}j_{0}^{'}  +  v_{th}\rho_{s}\frac{1}{r}\frac{\partial\left(\tilde{\psi,}\rho_{s}^{2}\nabla_{\perp}^{2}\tilde{\psi}\right)}{\partial\left(r,\theta\right)} + \nonumber\\ v_{th}\rho_{s}\left[\frac{1}{r}\frac{\partial\left(\tilde{W,}\tilde{\phi}\right)}{\partial\left(r,\theta\right)}+\left(\frac{n_{0}\left(0\right)}{2n_{0}}\right)\frac{1}{r}\frac{\partial\left(\tilde{W,}\tilde{n}\right)}{\partial\left(r,\theta\right)}\right]-   \frac{\rho_{s}^{2}W_{0}^{'}}{r}\frac{\partial\tilde{\phi}}{\partial\theta}+\nu\nabla_{\perp}^{2}\tilde{W}
\end{eqnarray}

\begin{eqnarray}
\label{Eq.3}
\fl \frac{\partial\tilde{\psi}}{\partial t}+\mathrm{\mathbf{v_{e0}}}\cdot\nabla\tilde{\psi}+v_{A}\nabla_{\parallel}\tilde{\phi}=v_{A}\left(\frac{n_{0}\left(0\right)T_{e0}}{n_{0}T^{*}}\right)\nabla_{\parallel}n^{*}+ \nonumber\\ \frac{v_{th}\rho_{s}}{r}\left[\frac{1}{r}\frac{\partial\left(\tilde{\psi,}\tilde{\phi}\right)}{\partial\left(r,\theta\right)}+\left(\frac{n_{0}\left(0\right)}{2n_{0}}\right)\frac{1}{r}\frac{\partial\left(\tilde{\psi,}\tilde{n}\right)}{\partial\left(r,\theta\right)}\right]+  \frac{c^{2}\eta}{4\pi}\nabla_{\perp}^{2}\tilde{\psi}
\end{eqnarray}

\begin{eqnarray}
\label{Eq.4}
\fl \frac{\partial\tilde{n}}{\partial t}+\mathrm{\mathbf{u_{e0}}}\cdot\nabla\tilde{n}+v_{A}\nabla_{\parallel}\rho_{s}^{2}\nabla_{\perp}^{2}\tilde{\psi}= \nonumber\\
v_{\rho s}\frac{1}{r}\frac{\partial\tilde{\psi}}{\partial\theta}\frac{4\pi\rho_{s}}{cB_{0}}j_{0}^{'}+v_{th}\rho_{s}\frac{1}{r}\frac{\partial\left(\tilde{\psi,}\rho_{s}^{2}\nabla_{\perp}^{2}\tilde{\psi}\right)}{\partial\left(r,\theta\right)}+ \nonumber\\ v_{th}\rho_{s}\frac{1}{r}\frac{\partial\left(\tilde{n,}\tilde{\phi}\right)}{\partial\left(r,\theta\right)} +v_{th}\rho_{s}\left(\frac{n_{0}^{'}}{N^{*}}\right)\frac{1}{r}\frac{\partial\tilde{\phi}}{\partial\theta}- v_{th}\nabla_{\parallel}\mathbb{{\normalcolor \tilde{\xi}}}+D\nabla_{\perp}^{2}n^{*}
\end{eqnarray}

\begin{eqnarray}
\label{Eq.5}
\fl \frac{\partial\tilde{\xi}}{\partial t}+\mathrm{\mathbf{u_{0}}}\cdot\nabla\tilde{\xi}+v_{th}\nabla_{\Vert}\tilde{n}= \nonumber \\
\left(\frac{n_{0}(r)v_{\Vert0}^{'}}{n_{0}(0)}\right)\rho_{s}\frac{1}{r}\frac{\partial\tilde{\phi}}{\partial\theta}+v_{th}\rho_{s}\frac{1}{r}\frac{\partial\left(\tilde{\xi,}\tilde{\phi}\right)}{\partial\left(r,\theta\right)}- \nonumber \\
v_{th}\rho_{s}\beta^{1/2}\left(\frac{p_{0}^{'}}{P^{*}}\right)\frac{1}{r}\frac{\partial\tilde{\psi}}{\partial\theta}-v_{th}\rho_{s}\beta^{1/2}\frac{1}{r}\frac{\partial\left(\tilde{p,}\tilde{\phi}\right)}{\partial\left(r,\theta\right)}- 
v_{th}\left(\frac{n_{0}}{N^{*}}\right)+\chi\nabla_{\perp}^{2}\tilde{\xi}
\end{eqnarray}

Equation[\ref{Eq.1}] is the Poisson relation for our system. Equation[\ref{Eq.2}] is the vorticity equation, where $\tilde{W}$ is the perturbed vorticity. Equation[\ref{Eq.3}] describes the evolution of the perturbed poloidal flux function $\tilde{\psi}$. Equation [\ref{Eq.4}] describes density evolution and equation [\ref{Eq.5}] describes the evolution of parallel momentum. Here, $\mathbf{u_{0}}=-\frac{cE_{r0}}{B}\mathbf{e_{\theta}}+\mathbf{b_{0}}v_{\Vert0}$ is the equilibrium `MHD' flow, and $\mathbf{u_{e0}}=-\frac{cE_{r0}}{B}\mathbf{e_{\theta}}+\mathbf{b_{0}}\left(v_{\Vert0}-j_{\Vert0}/en_{0}\right)$ is the corresponding electron flow. The ion flow alone is given by $\mathbf{v_{0}}=\mathbf{u_{0}}+\frac{c}{en_{0}B}T\frac{\partial n_{io}}{\partial r}\mathbf{e_{\theta}}$, here due to quasi-neutrality $n_{i0} \thicksim n_{e0}
 $ and $\mathbf{v_{e0}}=-\left[\frac{cE_{r0}}{B}+\frac{c}{en_{0}B}T\frac{\partial n_{e0}}{\partial r}\right]\mathbf{e_{\theta}}$ is the total electron poloidal flow composed of the electron $\mathbf{E} \times \mathbf{B}$  equilibrium flow and the electron diamagnetic flow.      Also, we have $N^{*}=n_{e}\left(0,t\right)$,   $T^{*}=T_{e}\left(0,t\right)+T_{i}\left(0,t\right)$,    $\tilde{\xi}=N^{*}V_{TH}$.  The resistivity $\eta$ and viscosity $\nu$ are specified quantities and are held constant during our calculations. In particular, we use the self-consistent formulation whereby $\eta(r)\j_{0z}(r) \equiv E_{0z} \equiv \frac{V_{loop}}{2\pi R_{0}}$, where the specified $q$ profile and $B_{0}$ are used to get $j_{0z}$ initial profile. After this, we hold the profile and the value of $\eta(0)$ fixed throughout both linear and nonlinear calculations. We also have, $D_{res}=\frac{c^2\eta(r)}{4\pi}$,$\nu \equiv D_{visc} \equiv Pr.D_{res}$, therefore,
\begin{equation}
Pr\equiv\frac{D_{visc}}{D_{res}}
\end{equation}

is the Prandtl number, which we have introduced earlier. We can therefore we see that kinematic viscosity $\nu$ and $\eta$ share the same radial profile and are invariant in time.

This comes from 
\[\frac{\delta\mathbf{E}}{B_{0}}=-\nabla\tilde{\phi}-\frac{1}{c}\frac{\partial\tilde{\phi}}{\partial t}\mathbf{e_{\zeta}}\] where, $\mathbf{E}$ is the electric field, $\phi$ is the electrostatic potential, and  $\mathbf{B_{0}\simeq} B_{0z}{\bf e}_{\zeta}+B_{0\theta}(\rho){\bf e}_{\theta}$  is the equilibrium field.  The fluctuating electric field, $\delta E$, is related to $\tilde{\phi}$ [this has dimensions of length].
We use, $\epsilon=a/R_{0}$, the inverse aspect ratio,  $v_{0}=V_{0z}\left(\rho\right)\mathbf{e}_{\zeta}+a\rho\Omega\left(\rho\right)\mathbf{e}_{\theta}$. 

 The magnetic field perturbations are normalized by the equilibrium axial magnetic field $B_{0z}$. The fluctuations of magnetic field and velocity are incompressible in the $(r- \theta)$ plane. The temperatures are measured in energy units, i.e, electron volts. 
Also, we have used fixed boundary conditions, along with a conducting boundary, which means that all the variables are zero at $\rho=1$. Additionally, regularity considerations mean that at $\rho=0$, the fluctuations approach zero, and we have set the plasma edge at $\rho=0.95$. We have used the Fourier representation for the purpose of periodicity of the angular coordinates.

Together, these equations constitute the four field model we use and we solve them using the CUTIE (\textbf{CU}lham \textbf{T}ransporter of \textbf{I}ons and \textbf{E}lectrons) code \cite{Thyagaraja2000, Chandra2015}, a nonlinear, global, electromagnetic, quasi-neutral, two fluid initial value code. It has been used earlier for studies of kink modes, tearing modes, ELMs, L to H transitions, internal transport barriers and other problems \cite{Thyagaraja2000,Thyagaraja2010,Chandra2015,Chandra2017,Mendonca2018}.  

We briefly describe the numerical details of our investigations. We have used a spatial resolution of 1801 radial grid points, 9 poloidal, 5 toroidal Fourier modes for the linear runs made with the resolvent code\cite{Chandra2015}, a version of CUTIE which is an eigenvalue solver, and is equivalent to the evolutionary version, and helps to find additional eigenvalues, as the evolution version only finds the fastest growing mode. In the nonlinear case, we reduce the radial grid to 101 points, due to limited computational resources. In the nonlinear cases, the error in growth rate thus introduced is not significant as we are more concerned with nonlinear saturated energy levels, a high accuracy in growth rate to many decimal places does not yield any additional information given the approximations in our model.   

\section{Results of linear simulations}\label{linear results}
%

\subsection{No flow simulations}

In this section we report on simulations without an imposed flow , $i.e.$, $v_{0\theta} \equiv v_{0\zeta} \equiv 0$. Our particular emphasis is to study the variation of the mode characteristics as a function of Pr. In our simulations we have calculated the diamagnetic frequency $\omega*$ using the following formula:

\begin{equation}
\omega*=-\frac{cT}{eB_{0}}\left[\frac{1}{n_{e}}\right]\frac{dn_{e}}{dr}\left[\frac{m}{r}\right]\left(rad/s\right)
\end{equation} 

Here, $T=275 eV$, $B=2\times 10^4 gauss$, $e=4.8 \times 10^{-10} $ stat.coul, $c=3 \times 10^{10} $ cm/s $\frac{cT}{eB}=1.37\times10^6  cm^2/s,\frac{1}{n_e}\frac{dn_{e}}{dr}=-2\alpha\left(r/a^2\right) $. Substituting these values, we obtain,

\begin{equation}\label{omega_eqn}
\omega*\tau_{A}=1.255 \times \alpha \times 10^{-3}
\end{equation}

We note the values of the following parameters used in our runs: $v_{A}=2.18 \times 10^{8}$ cm/s, $ \beta = 1.6 \%$,$\tau_{A}= 4.58 \times 10^{-8}$,$\tau_{\eta}=4.58 \times 10^{-2}$ and $\tau_{\nu}= Pr \times \tau_{\eta}$, the value of $Pr$ is specified where used. In the following Fig. \ref{fig:d_res_profile} we have showed a typical profile for $D_{res}$, and since $D_{visc}=Pr \times D_{res}$, thus it has the same profile because $Pr$ is constant in our model.

\begin{center}
\begin{figure}[!htb]
\centering
\includegraphics[scale=0.3]{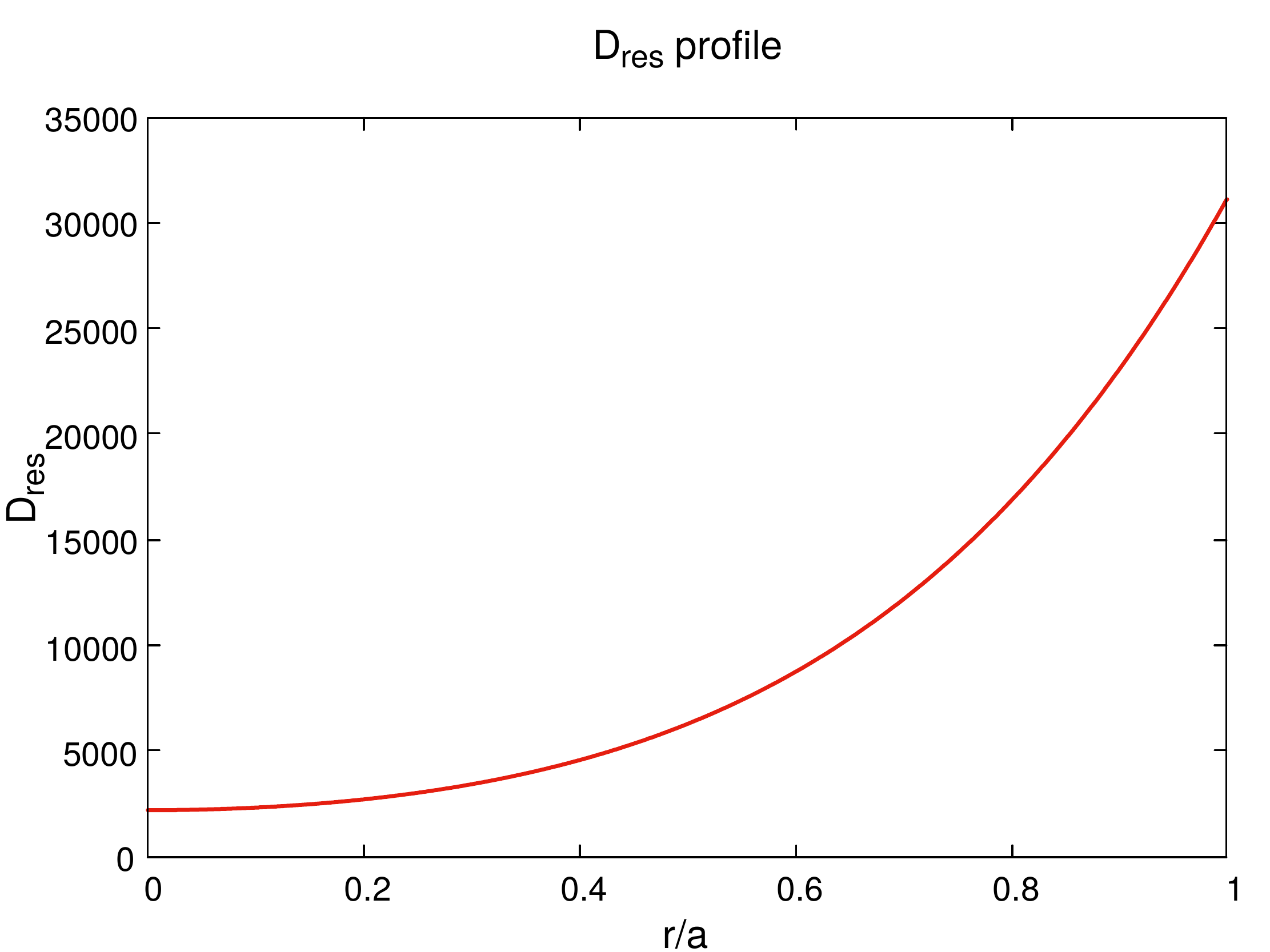}    
\caption{$D_{res}$ profile}
\label{fig:d_res_profile}
\end{figure}
\end{center}
 
The eigenvalue version of CUTIE, called Resolvent-CUTIE was used to compute these linear results\cite{Chandra2015}.
 
\begin{center}
\begin{figure}[!htb]
\centering
\includegraphics[scale=0.35]{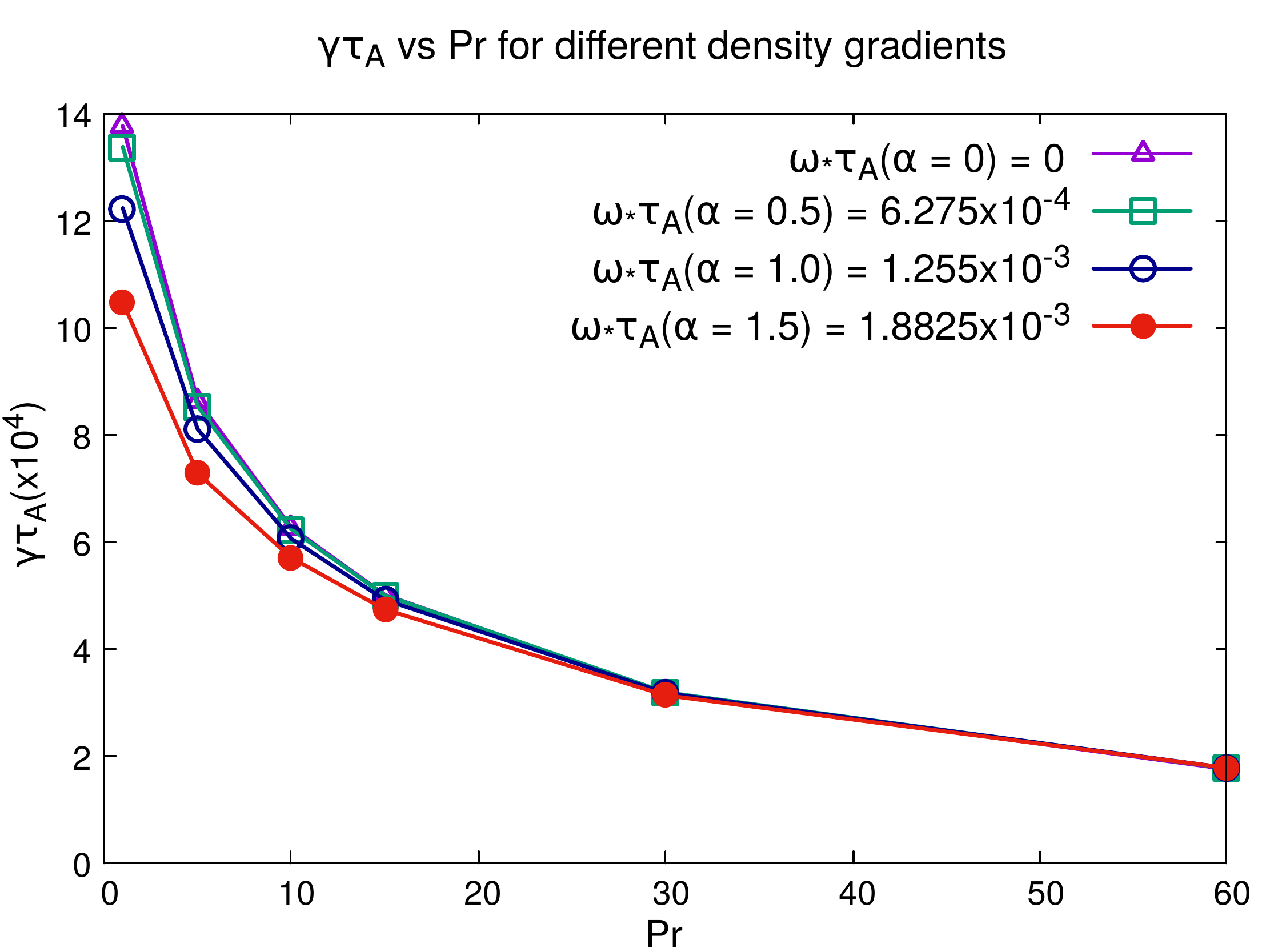}     
\caption{Growth rate variation with Pr without flow}
\label{fig:omega_star_growth}
\end{figure}
\end{center}

\begin{center}
\begin{figure}[!htb]
\centering
\includegraphics[scale=0.35]{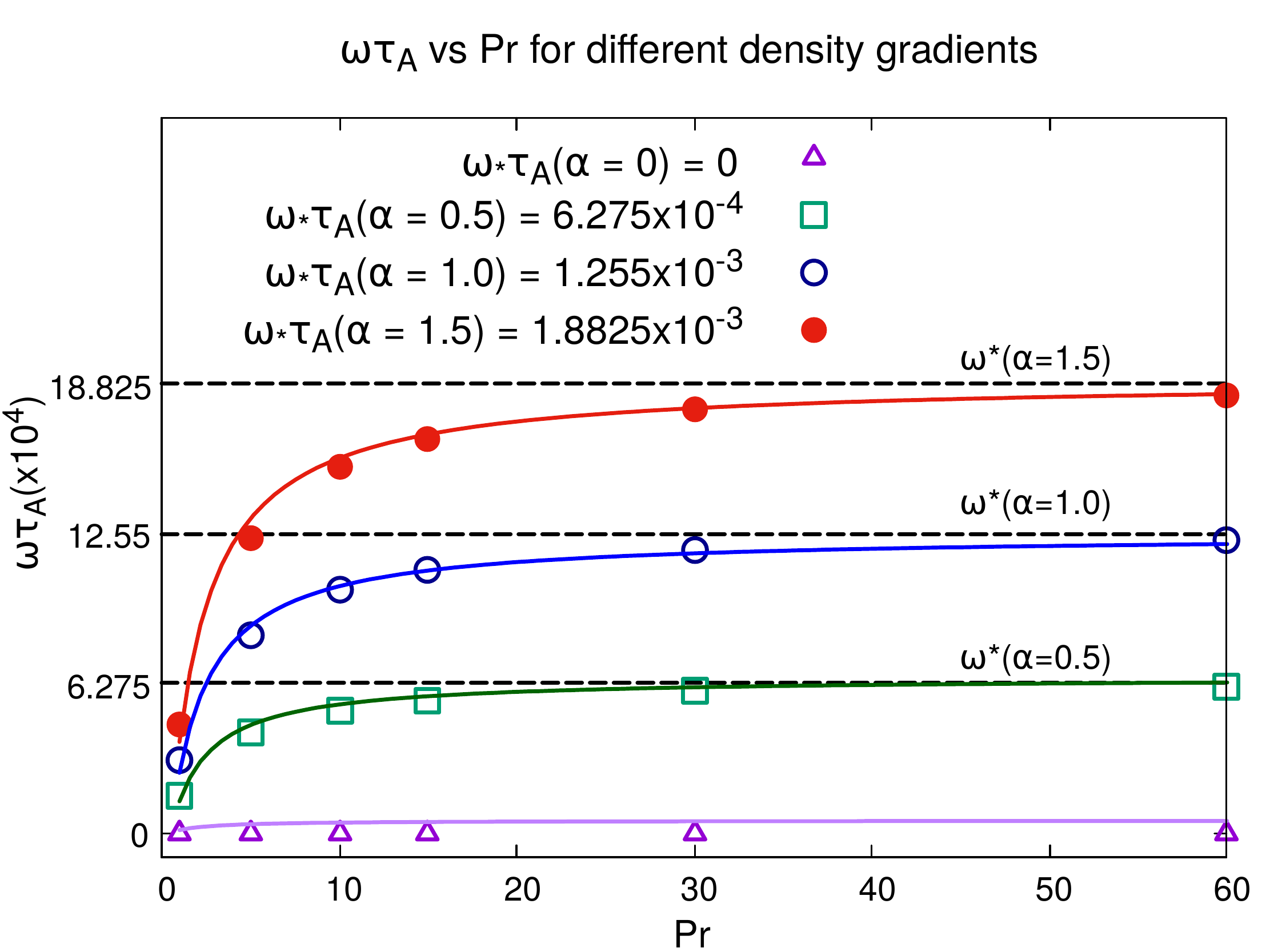}     
\caption{Frequency variation with Pr without flow . The points for each $\alpha$ value are fitted with a curve using the formula $\omega\tau_{A}\propto\frac{(Pr-0.5)}{Pr+1.5}$}
\label{fig:omega_star_frequency}
\end{figure}
\end{center}

Our results are plotted in Fig. [\ref{fig:omega_star_growth}] and Fig .[\ref{fig:omega_star_frequency}].

In figure [\ref{fig:omega_star_growth}], we notice that the growth rate reduces smoothly with Prandtl number, Pr, in accordance with the expectation that higher viscosity provides a larger damping.

 In figure [\ref{fig:omega_star_frequency}] we see that the real frequency of the mode increases as a function of $Pr$ and approaches the $\omega*$ frequency asymptotically. This is a novel result showing that viscosity not only influences the growth rate (as shown in figure \ref{fig:omega_star_frequency}) but it can also influence the real frequency. This reactive contribution from viscosity was earlier noticed (but not commented upon) by Porcelli et al\cite{Porcelli-Migliulo1986} over a small range of $Pr$. Our simulations establish  the existence of such a contribution over a wide range of $Pr$ values. Interestingly, the nature of this variation of the real frequency as a function of Pr can be modeled by the following relation, $\omega\tau_{A}\propto\frac{(Pr-0.5)}{Pr+1.5}$, where the proportionality constant changes with the $\alpha$ value.  This scaling relation is shown as solid curves in 
 Fig.~ \ref{fig:omega_star_frequency} and as can be seen it relation provides a good analytic fit to the data points for all values of $\alpha$.  We also note that the growth rate is almost independent of $\alpha$ once Pr exceeds 15, as the viscous diffusion layer then becomes larger than the resistive layer.

\subsection{Simulations with imposed flow} \label{linear flow results}

\subsubsection{Axial Flows}
We have used an  imposed axial flow profile of the form 

\begin{equation}
V_{0z}/V_{A}=M_{z}tanh(\rho -\rho_{s})
\end{equation} 

where, $V_{0z}$ is the imposed axial flow velocity, $V_{A}$ is the Alfv\'{e}n velocity, $M_{z}$ is the axial Mach Number, $\rho=\frac{r}{a}$, is the normalized radial coordinate and $\rho_{s}=\rho$ when $r=r_{s}$ is the resonant radius. In this section and henceforth we refer to the $q=1$ resonant radius as $\rho_{q=1}$.

The following Fig. \ref{fig:axial_flow_profile} is a typical profile for $V_{0z}$,

\begin{center}
\begin{figure}[!htb]
\centering
\includegraphics[scale=0.35]{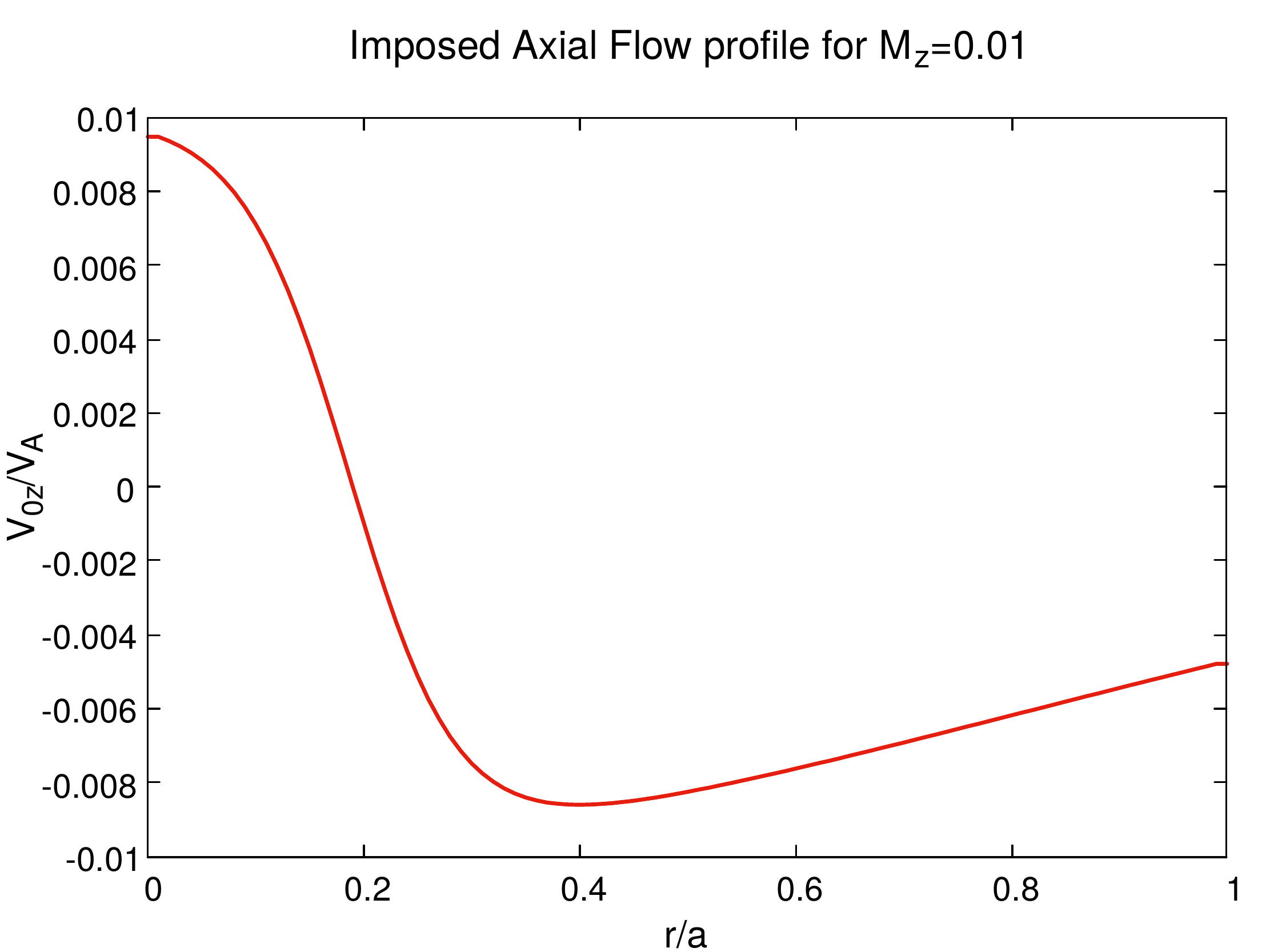} 
\caption{Imposed axial flow profile}
\label{fig:axial_flow_profile}
\end{figure}
\end{center}

This profile is chosen to have the property of zero flow at the resonant surface, so we can study the effect of shear alone at the resonant surface on the mode. We state the values of $\omega_{*}(\alpha)\tau_{A}$ as a function of $\alpha$ as given in Eqn. \ref{omega_eqn} for convenience:

\begin{table}[!htbp]
\caption{Value of $\omega_{*}(\alpha)\tau_{A}$ }
\centering
\begin{tabular}{|c|c|}
\hline 
\multicolumn{1}{|c|}{$\alpha$} & \multicolumn{1}{|c|}{$\omega_{*}(\alpha)\tau_{A}$ $\times 10^{-4}$}\\
\hline
0.5  &  6.275  \\  
1.0  &   12.25  \\  
1.5  &   18.82  \\
\hline  
\end{tabular} 
\end{table}

\begin{center}
\begin{figure}[!htbp]
\centering
\includegraphics[scale=0.35]{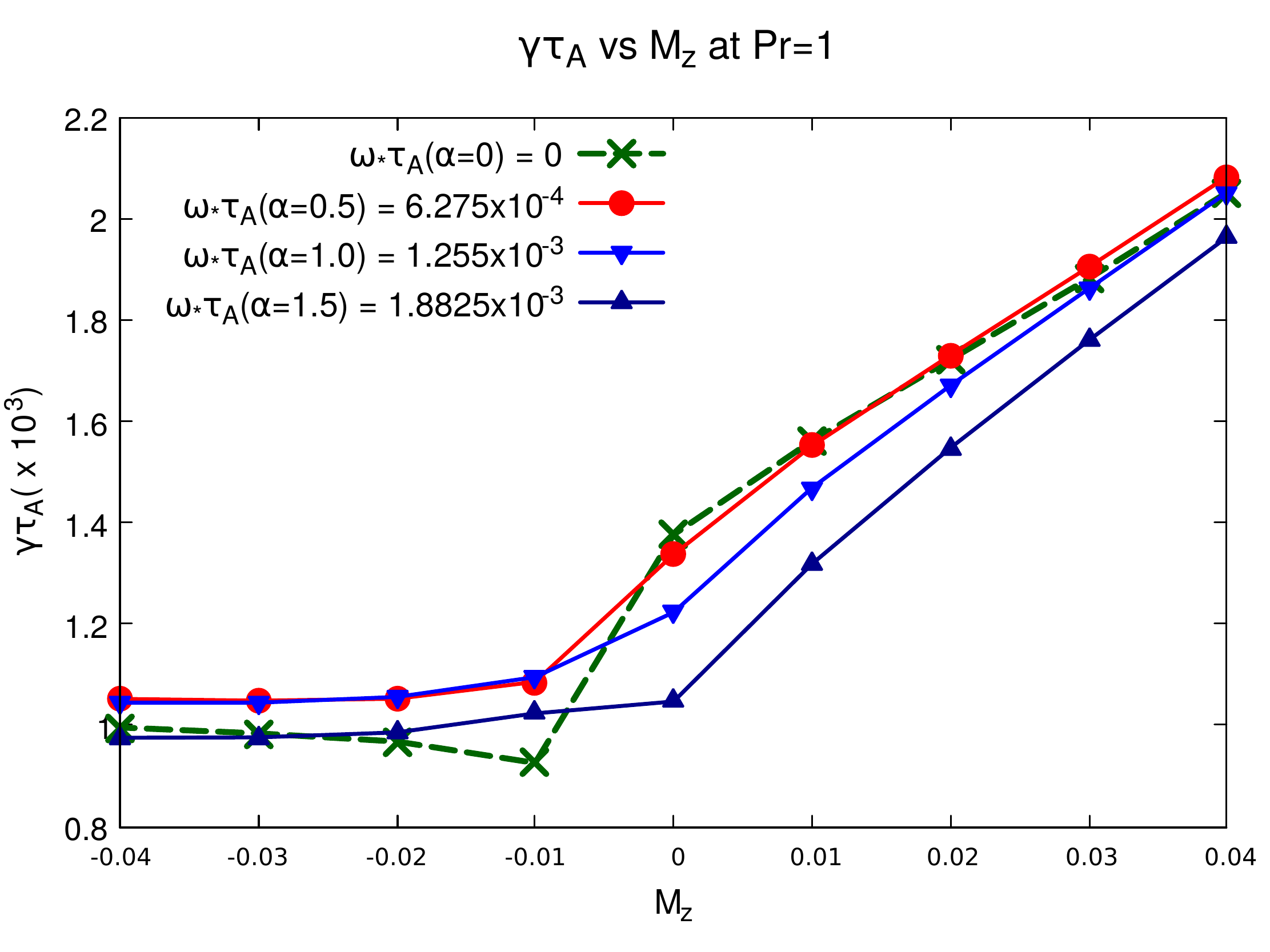} 
\caption{\small Normalized growth rate vs Axial Flow(Axial Mach Number) for different density gradients at a fixed Pr=1}
\label{axgr1}
\end{figure}
\end{center}

We have studied the variation in growth rate and real frequency of the internal kink mode in two ways. We have studied the variation of the growth rate  for different $\alpha$ values, in Fig. \ref{axgr1}. In the Fig. \ref{axgr1}, we observe that the growth curves exhibit a very interesting behaviour, in that, for a pure RMHD case, we obtain symmetry of the normalized growth rate curve about $M_{z}=0$, that is changing the sign of $M_{z}$ does not change the growth rate, but here we see a distinct asymmetry. This is due to the contributions of the parallel momentum and electron continuity equations. We have done a study by eliminating the parallel momentum equations, and setting $\alpha=0$; in this case, we recover RMHD behaviour.

\begin{center}
\begin{figure}[!htbp]
\centering
\includegraphics[scale=0.35]{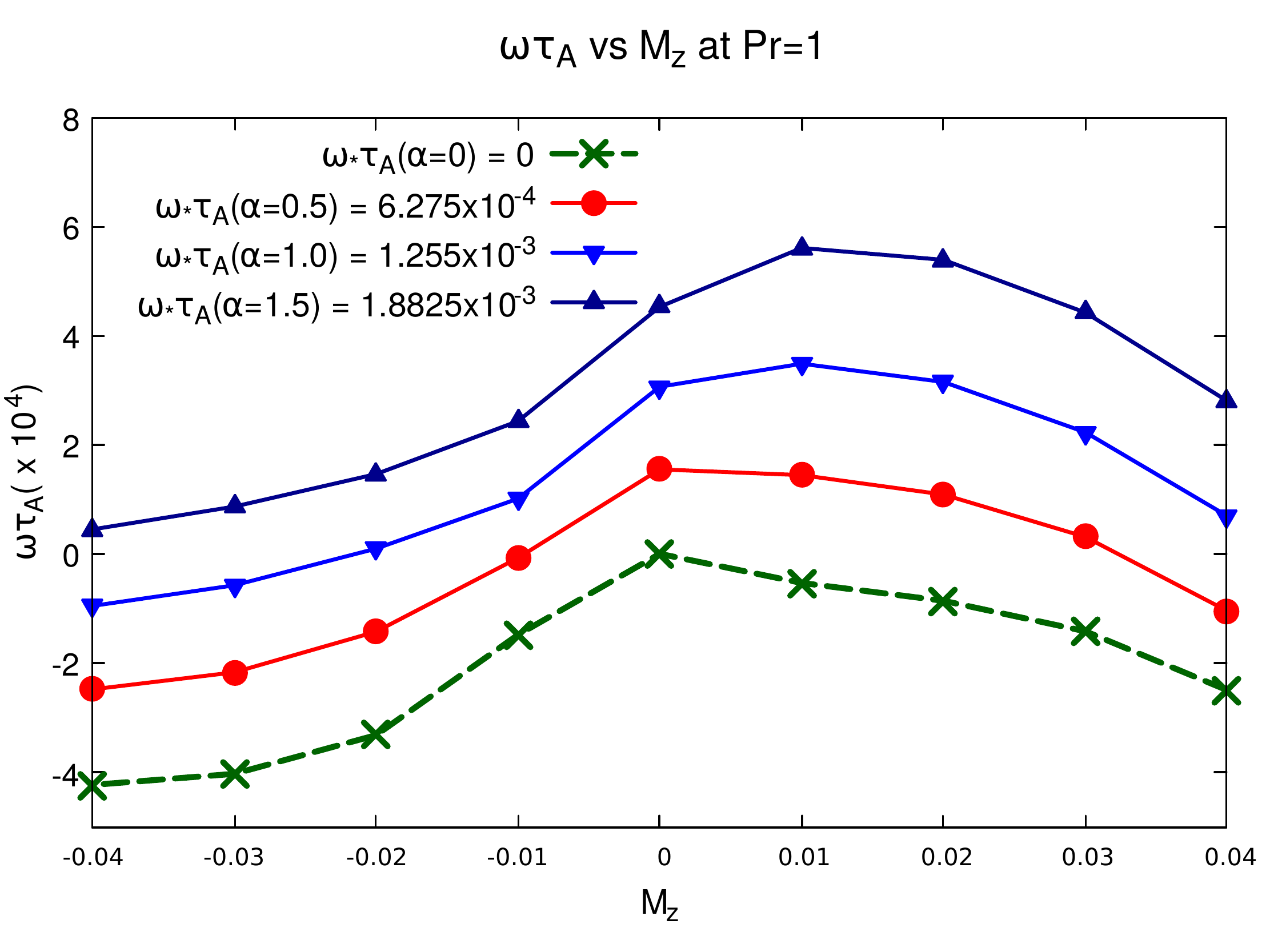} 
\caption{\small Frequency vs Axial Flow(Axial Mach Number) for different density gradients at a fixed Pr=1}
\label{axfr1}
\end{figure}
\end{center}

In a similar manner, in Fig. \ref{axfr1}, which shows the corresponding variation of real frequency with $M_{z}$ for different values of $\alpha$, we observe a similar asymmetry of frequency about $M_{z}=0$. This can be attributed to $\omega_{*}$-related contributions to the frequency of the mode, apart from the rotation caused by the axial flow. However, here all the curves are approximately parallel to each other, unlike in the growth rate case.

\begin{center}
\begin{figure}[!htbp]
\centering
\includegraphics[scale=0.35]{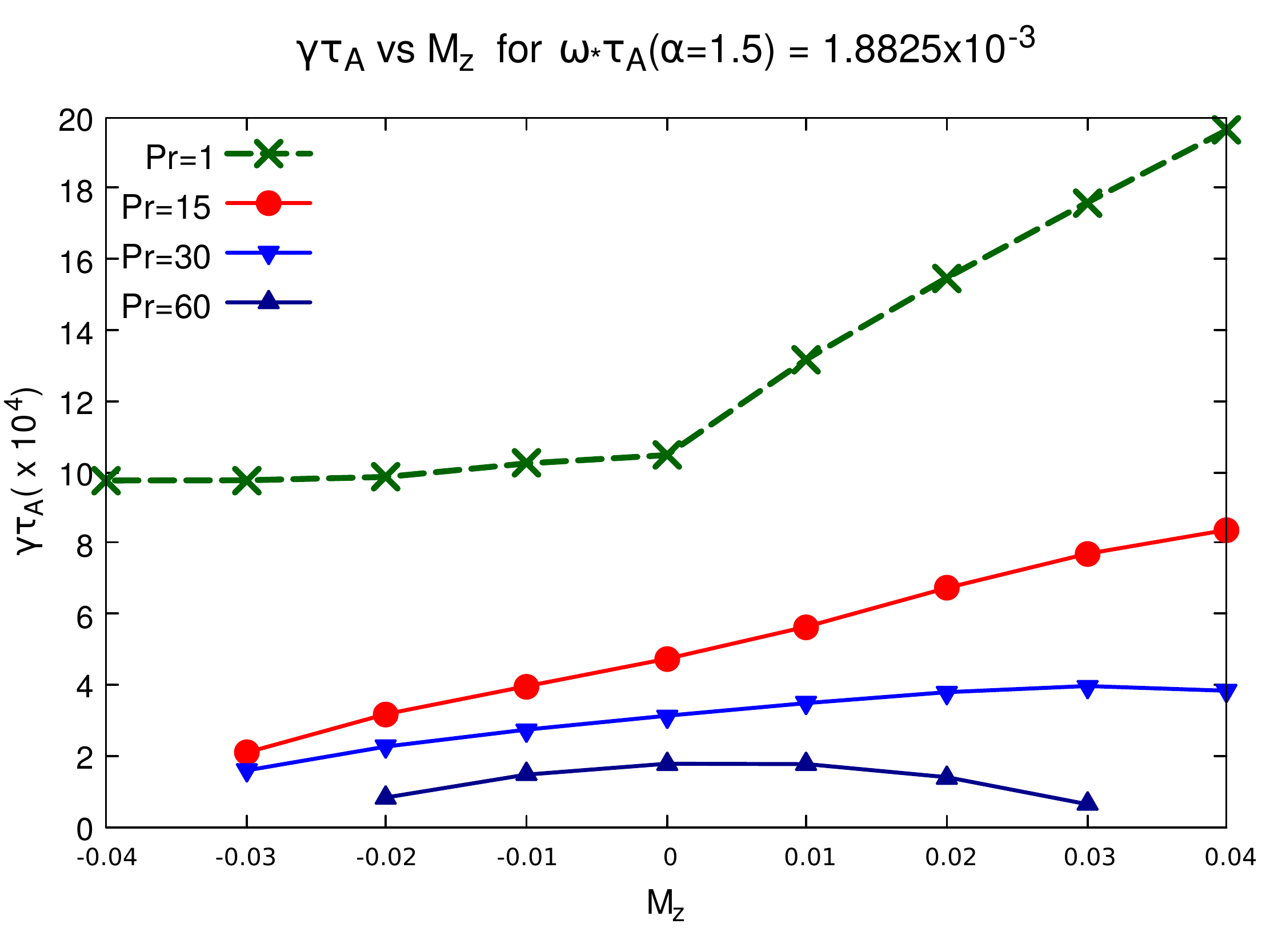} 
\caption{\small Normalized growth rate vs Axial Flow(Axial Mach Number) for different Pr at a fixed density gradient}
\label{axgr2}
\end{figure}
\end{center}

In Fig. \ref{axgr2} we plot the mode growth rate as a function of the axial flow Mach number, $M_{0z}$, for a given $\omega_{*}(\alpha=1.5)=1.8825\times10^{-3}$ for different $Pr$ values, at a fixed $\alpha$. In the Fig. \ref{axgr2}, the $\omega*\tau_{A}$ is fixed and corresponds to $\alpha=1.5$. It is seen that the curves are qualitatively different for $Pr=15$ and higher, and upon further investigation, we have found that the behavior seen for $Pr=1$ continues upto $Pr=8$, and after that the curve changes its shape. This suggests a transition in the behavior of the mode with respect to the viscosity (i.e. $Pr$ alone is changed) when imposed flows exist.

\begin{center}
\begin{figure}[!htbp]
\centering
\includegraphics[scale=0.35]{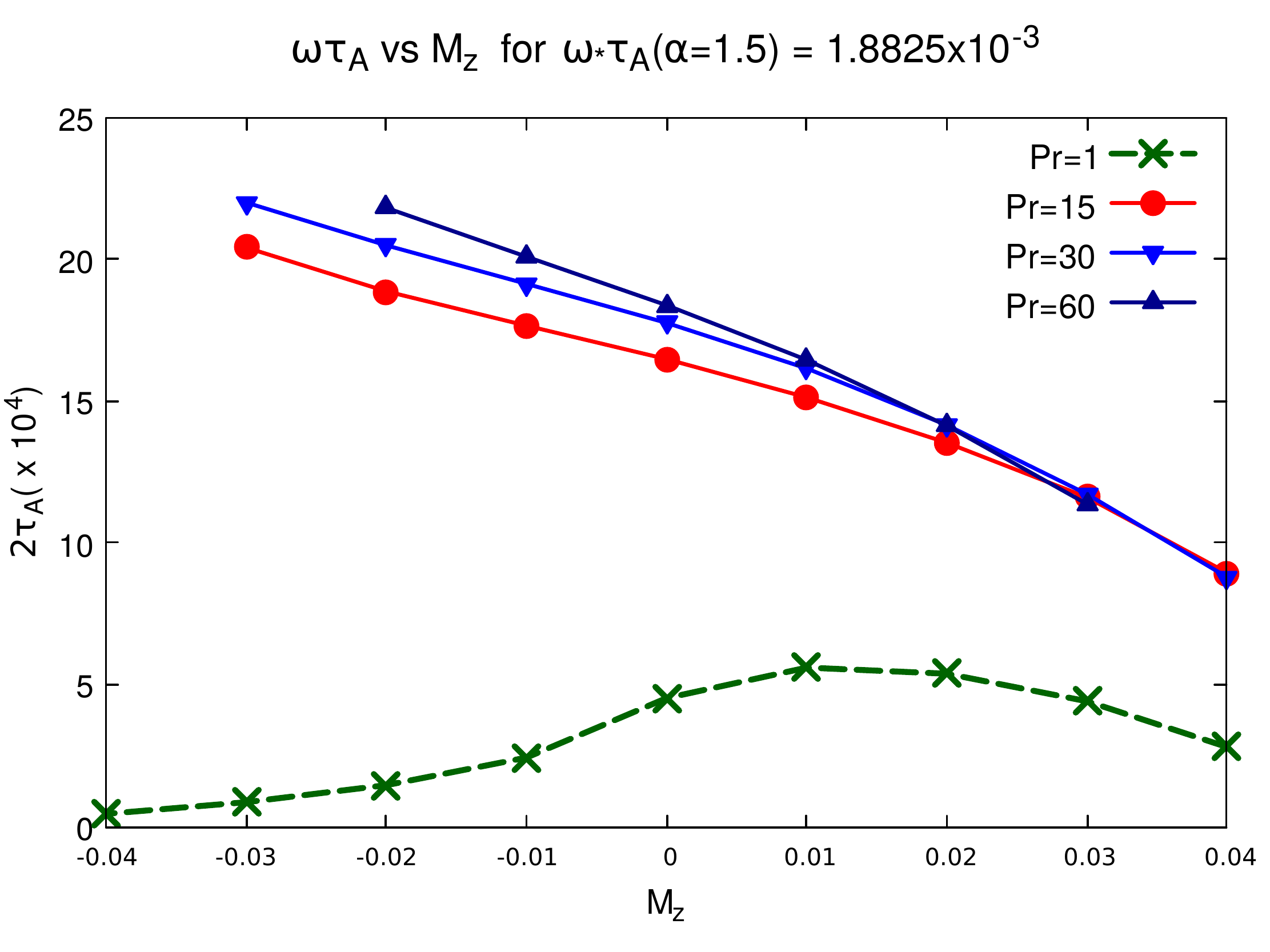} 
\caption{\small Frequency vs Axial Flow(Axial Mach Number) for different Pr at a fixed density gradient}
\label{axfr2}
\end{figure}
\end{center}

In Fig. \ref{axfr2}, we plot the real frequencies corresponding to the growth rates shown in Fig. \ref{axgr2}. It is clearly seen that, as before, for $Pr=1$, frequency variation with $M_{z}$ is different from those for $Pr=15$. It is also seen that the real frequencies increase with Pr.

In summary, we see a very strong asymmetry in the growth rates as a function of the sign of $M_{z}$.This is because a self consistent poloidal flow is generated by the diamagnetic drift. The frequencies show a similar profile as a function of $M_{z}$, which is unlike what we observe for the growth rates. This is probably due to the fact that the $\omega_{*}$ is independent of the axial flow and depends only on the density gradient in the linear regime.  We then have the case of varying Pr for a fixed, finite density gradient. In this case, we observe that while low Pr increases the growth rate, high Pr reduces it, consistent with our observations in the RMHD case. The shape of the curves for the $Pr=1$ case surprisingly is very different than for the other Pr's, which seems to indicate a threshold in Prandtl number after which the behavior of the mode changes. The frequency also shows similar trends, and it seems to be the case that frequency and viscosity are not directly proportional, contrary to the expectation that viscosity being dissipative in nature should gradually reduce the frequency, but in fact increases it up to a certain limit. 


\subsubsection{Poloidal flow}

We have used the following form for an imposed purely poloidal flow profile,
 
\begin{equation}
\frac{V_{0\theta}}{v_{A}}=M_{\theta}(\rho)
\end{equation}
 
where the poloidal Mach number,
\[M_{\theta}(\rho)=\rho\frac{a\Omega(\rho)}{v_{A}}\]

and \[\Omega(\rho)=\left(1-\rho^2\right)^{2}\]
 
Here, $V_{0\theta}$ is the equilibrium poloidal flow, and $\Omega(\rho)$ is the
poloidal angular frequency at $\rho=r/a$. The following Fig. \ref{fig:pol_flow_profile} is a typical flow profile:

\begin{center}
\begin{figure}[!htb]
\centering
\includegraphics[scale=0.35]{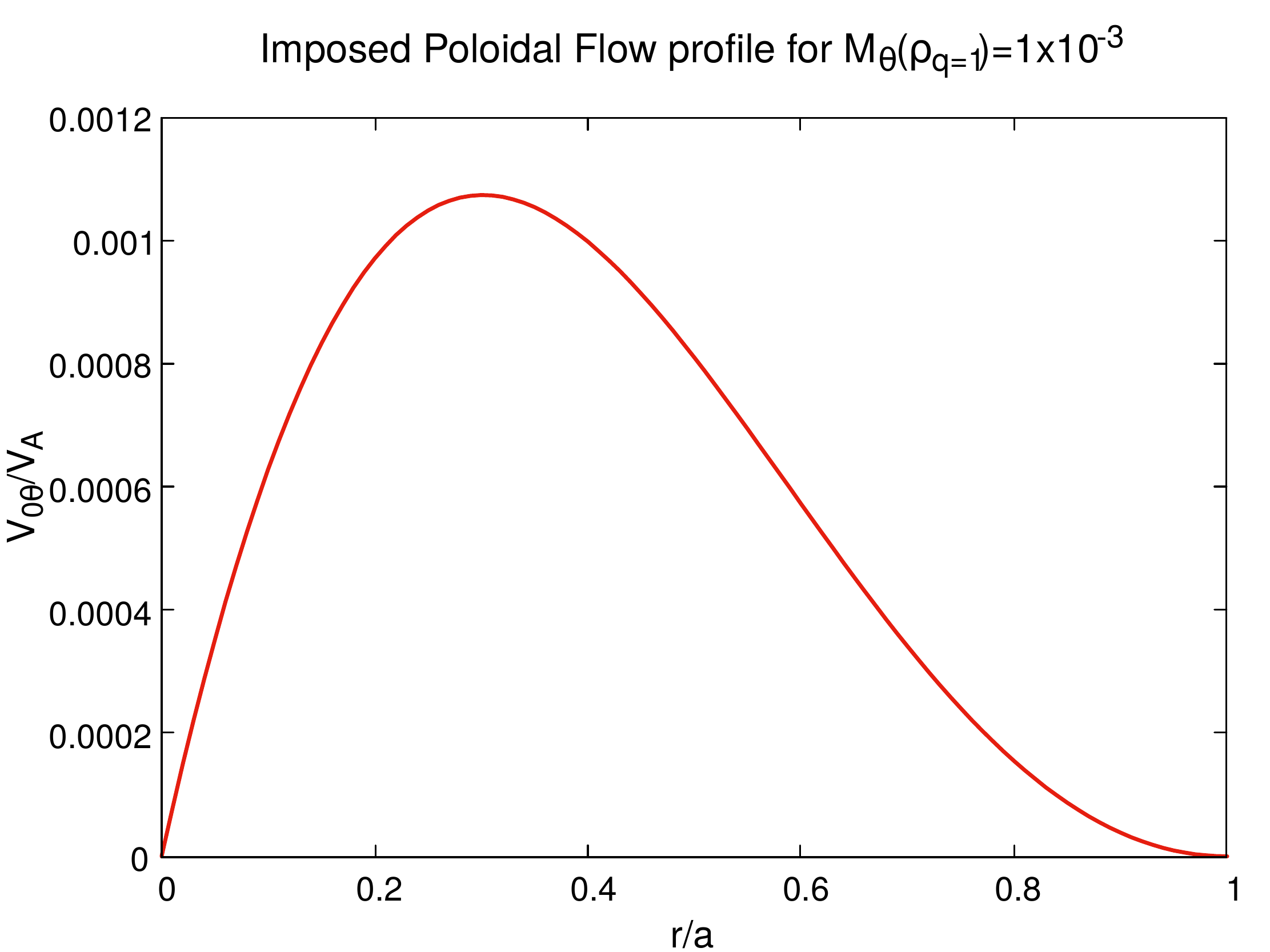} 
\caption{Imposed poloidal flow profile}
\label{fig:pol_flow_profile}
\end{figure}
\end{center}

\begin{figure}[!htbp]
\centering
  \begin{minipage}[h]{0.45\textwidth}
     \includegraphics[width=\textwidth]{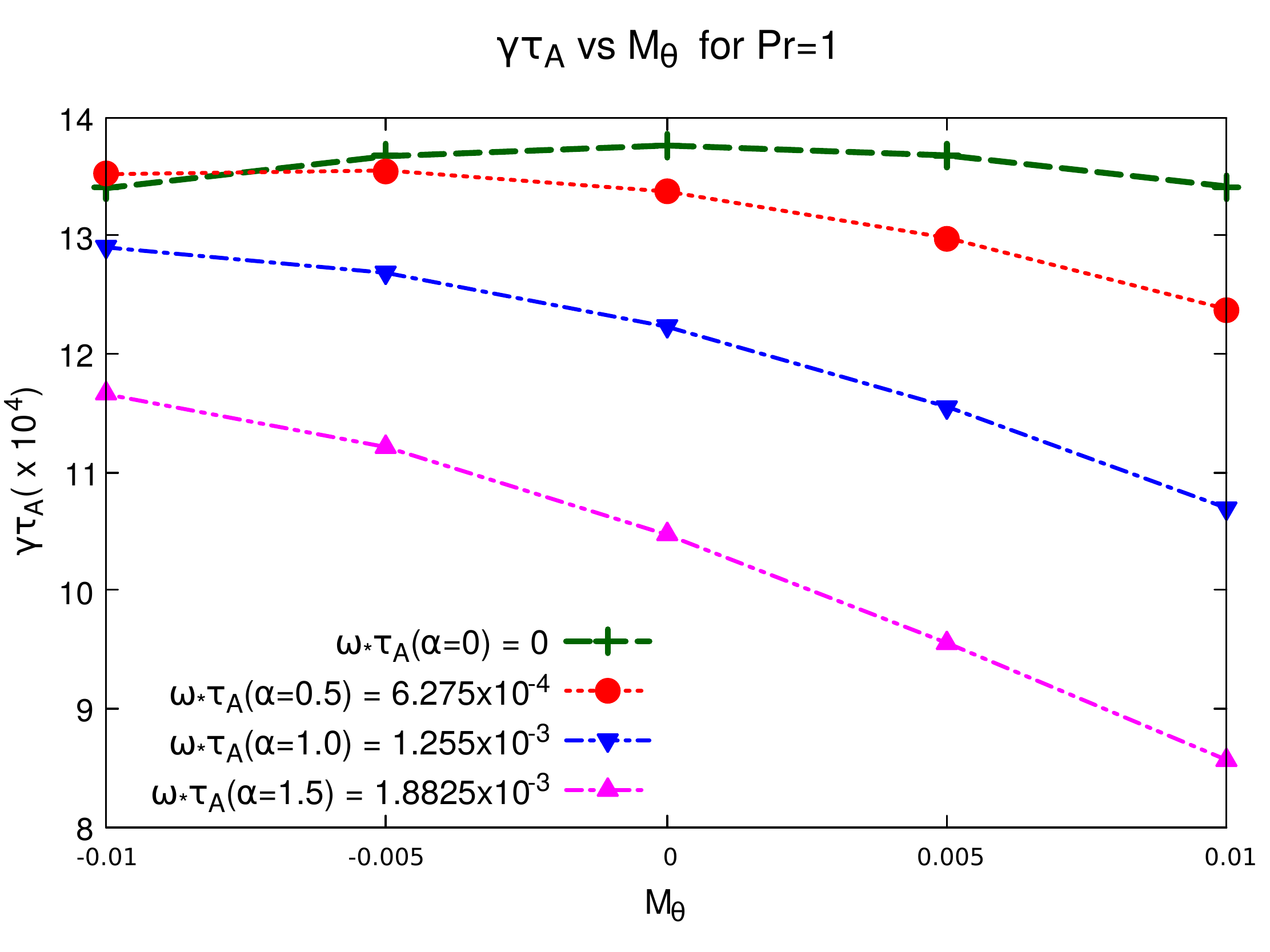} 
       \caption{\small Normalized growth rate vs Poloidal Flow($M_{\theta}\left(\rho_{q=1}\right)$) for different density gradients at a fixed Pr=1}
          \label{polgr1}  
  \end{minipage}
  \hfill
  \begin{minipage}[h]{0.45\textwidth}
      \includegraphics[width=\textwidth]{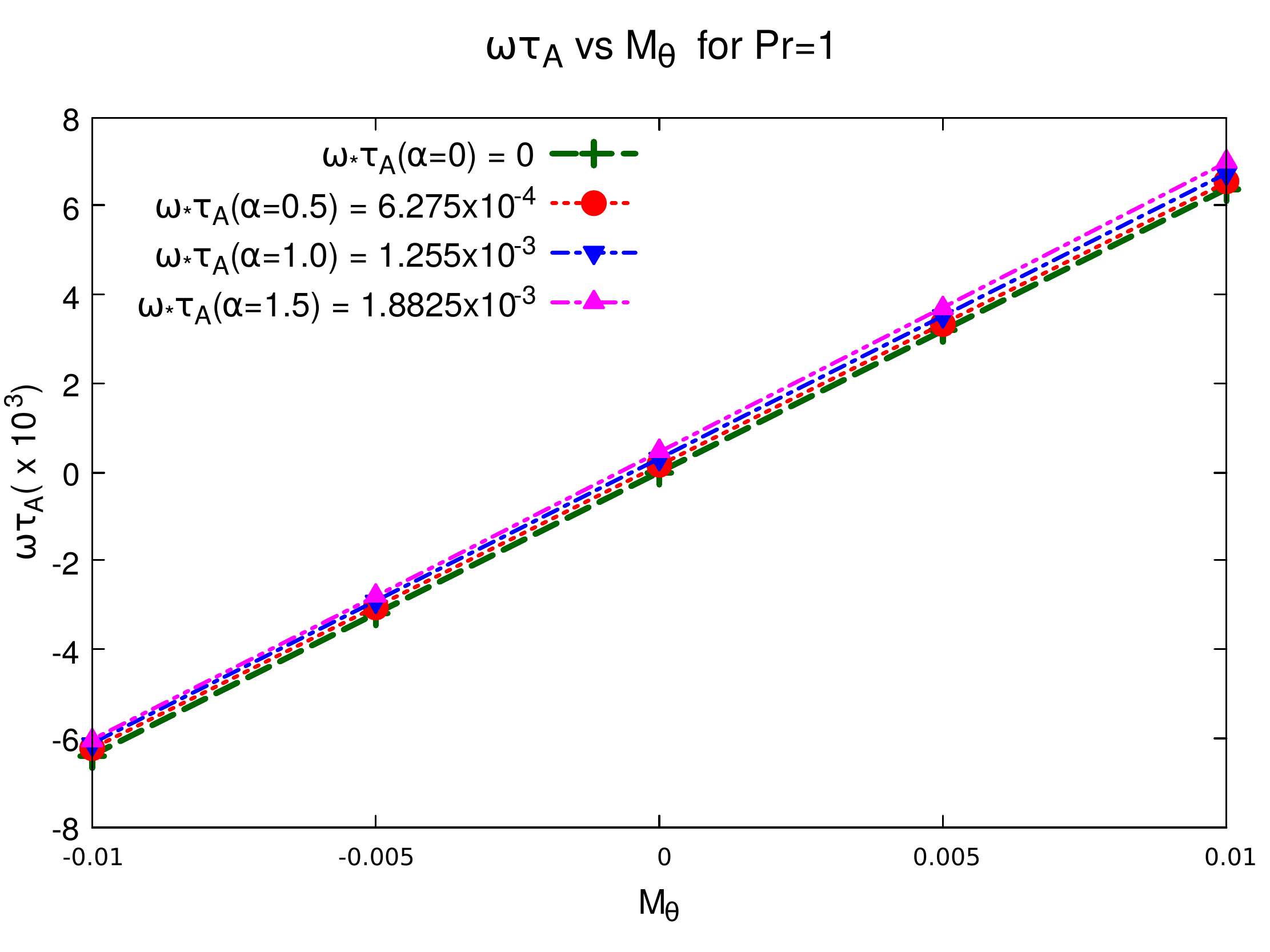} 
     \caption{\small Frequency vs Poloidal Flow($M_{\theta}\left(\rho_{q=1}\right)$) for different density gradients at a fixed Pr=1 }            
            \label{polfr1}
  \end{minipage}
\end{figure}

In Fig. \ref{polgr1}, we have plotted the normalized mode growth rate vs poloidal flow for different density gradients at a fixed Pr. In Fig. \ref{polgr1}, we see that poloidal flow is stabilising the mode with increasing $M_{\theta}\left(\rho_{q=1}\right)$ and decreases it in the negative direction. This is consistent with our finding that there is a positive $\omega_{*}\tau_{A}$ present in the system, thus it adds when the flow is positive, stabilising the mode, and cancels when the flow is negative, so the mode gets destabilised. We see further evidence of this in the $\alpha=0$ curve which is symmetric when the sign of $M_{\theta}\left(\rho_{q=1}\right)$ is changed.

Our hypothesis is further strengthened by observing Fig. \ref{polfr1} where we see that the frequencies vary linearly with $M_{\theta}\left(\rho_{q=1}\right)$, and the curves for different $\omega_{*}\tau_{A}$, i.e., $\alpha$ values are parallel to each other. 

\begin{figure}[!htbp]
\centering
  \begin{minipage}[h]{0.45\textwidth}
     \includegraphics[width=\textwidth]{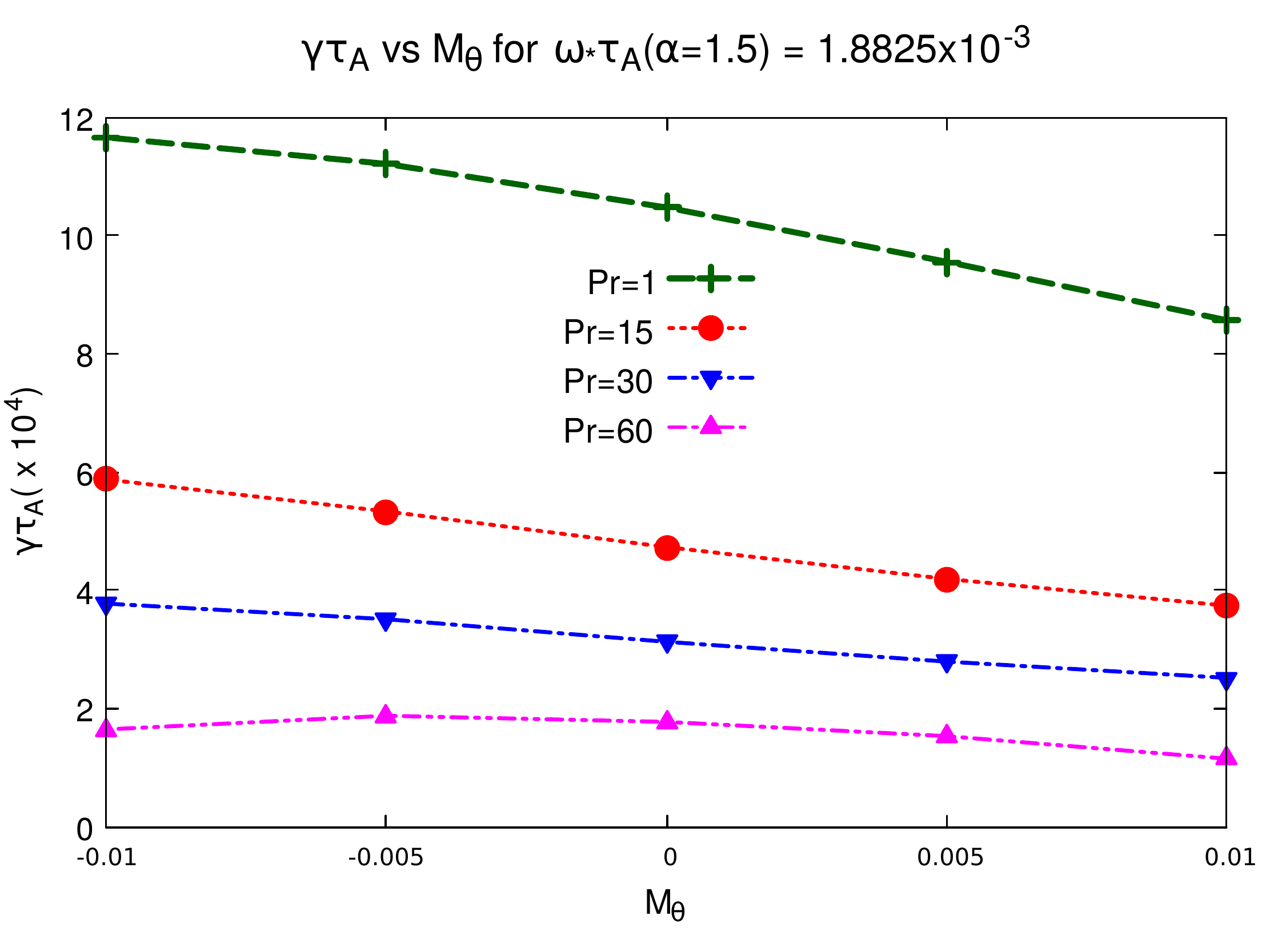} 
           \caption{\small Normalized growth rate vs Poloidal Flow($M_{\theta}\left(\rho_{q=1}\right)$) for different Pr at a fixed density gradient}
            \label{polgr2}
  \end{minipage}
  \hfill
  \begin{minipage}[h]{0.45\textwidth}
    \includegraphics[width=\textwidth]{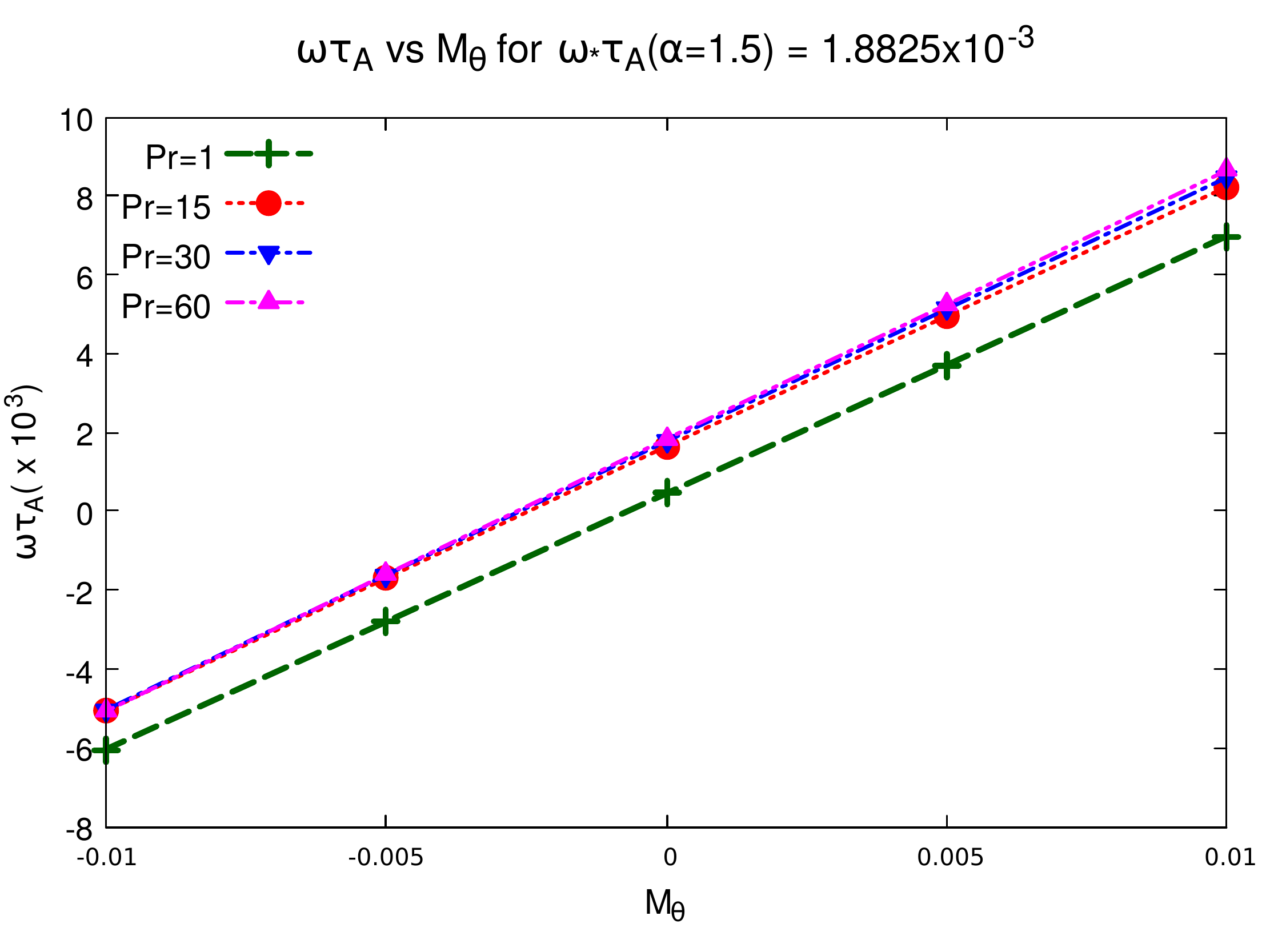} 
       \caption{\small Frequency vs Poloidal Flow($M_{\theta}\left(\rho_{q=1}\right)$) for different Pr at a fixed density gradient}
     \label{polfr2}    
   \end{minipage}
\end{figure}

In Fig.  \ref{polgr2}, where we have plotted normalized growth rates vs $M_{\theta}\left(\rho_{q=1}\right)$ for different Pr at a fixed $\alpha$. We notice that the trend for $Pr=1$ in Fig. \ref{polgr2} is qualitatively different from the other curves, but not substantially so. In the case of the corresponding frequency curve in Fig. \ref{polfr2}, we notice similarly that the $Pr=1$ curve is parallel to the other curves, which are almost overlapping. This is also consistent with our Fig. \ref{fig:omega_star_frequency}, where the frequency actually rises with Pr.  

In conclusion, we notice that the normalized growth rates are symmetric in the case of $\alpha=0$ as a function of the sign of $M_{\theta}$, which is consistent with what we see in the RMHD case. After we introduce a finite $\alpha$, we see curves which are broadly parallel, indicating a fixed $\omega_{*}$ in the system. Importantly, therefore, unlike in the axial flow case, the system is symmetric as a function of poloidal flow. The frequencies confirm this, being linear functions of $M_{\theta}\left(\rho_{q=1}\right)$ and are parallel for different values of $\alpha$. When we vary Pr for a fixed $\alpha$, the curves display broadly the same shape, indicating that Pr does not affect poloidal flow in an asymmetric fashion as it did in the case of axial flow. The frequencies in this case show an expected increase in frequency with Pr as we observed in the benchmarking section. 

\subsubsection{Helical Flow}

\begin{figure}[!htbp]
\includegraphics[scale=0.32]{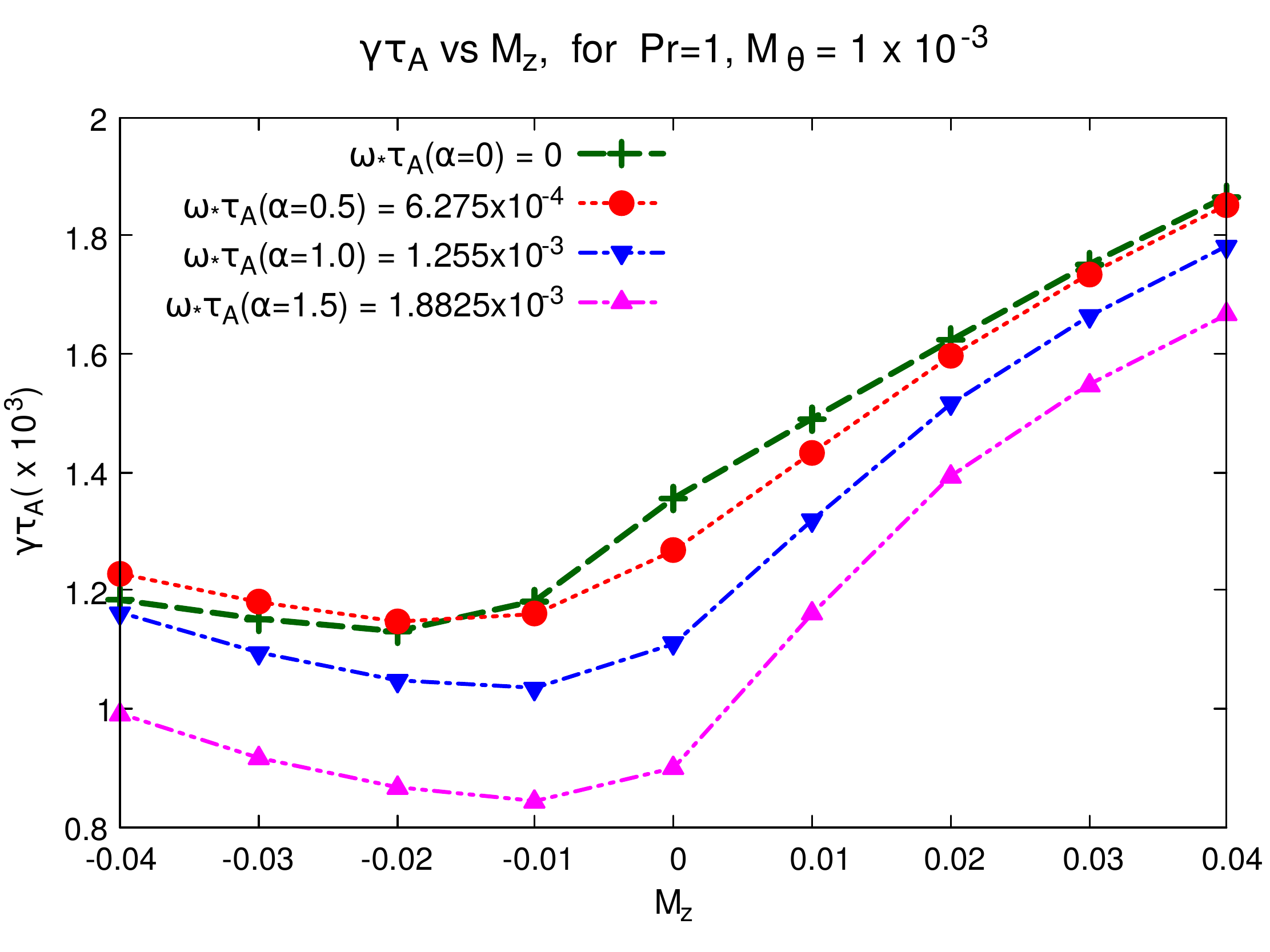} 
    \caption{\small Normalized growth rate vs Helical Flow($M_{z}$,fixed $M_{\theta}\left(\rho_{q=1}\right)=1\times10^{-3}$) for different density gradients at a fixed Pr=1}
      \label{helgr1}
\end{figure}

\begin{figure}[!htbp]
\includegraphics[scale=0.32]{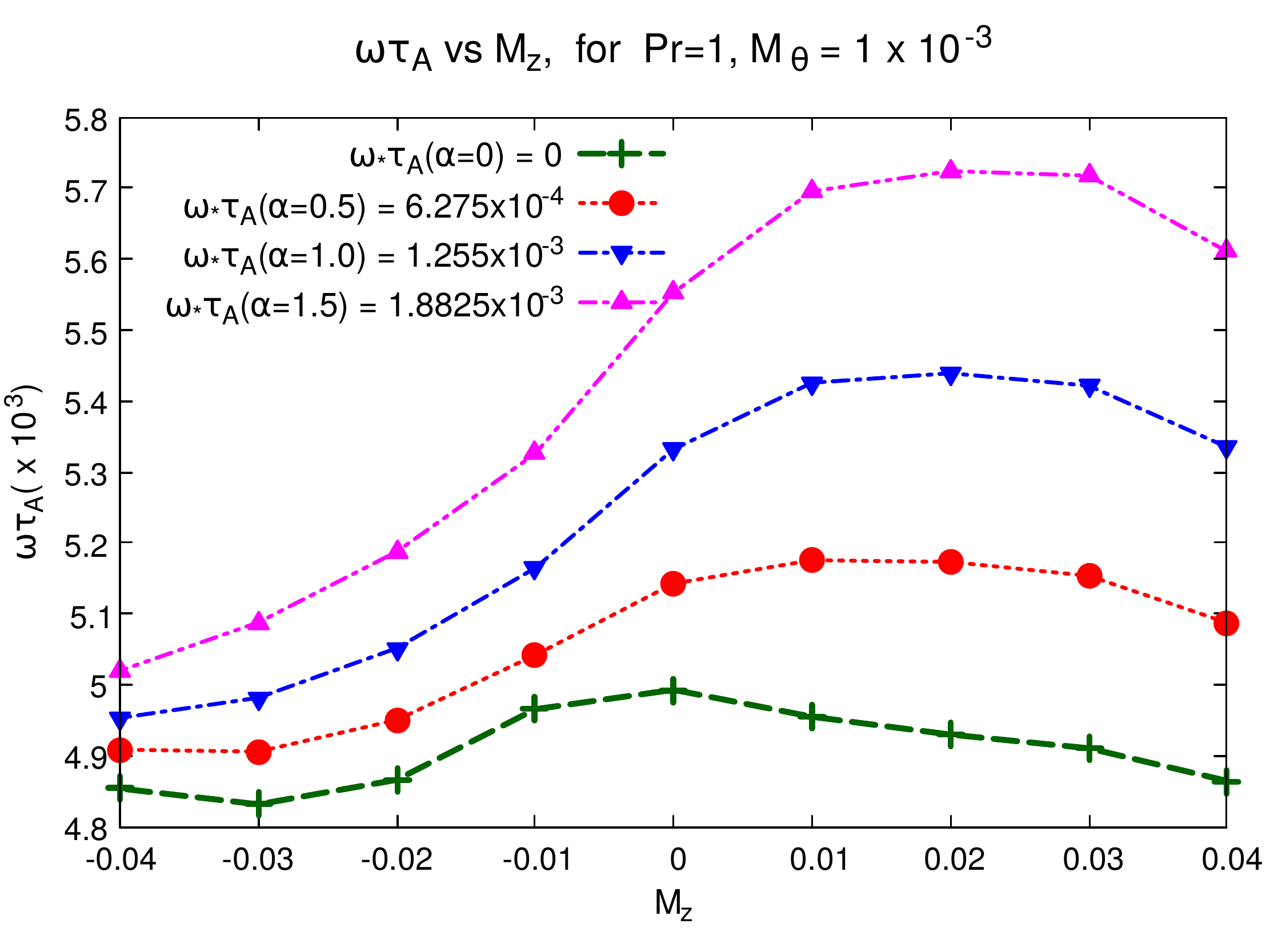}  
     \caption{\small Frequency vs Helical Flow($M_{z}$,fixed $M_{\theta}\left(\rho_{q=1}\right)=1\times10^{-3}$) for different density gradients at a fixed Pr=1}
      \label{helfr1}
\end{figure}

In this section we discuss helical flow results. Helical flows are a combined effect of axial and poloidal flows. We have used a fixed poloidal flow of $M_{\theta}\left(\rho_{q=1}\right)$, without loss of generality as our results are similar when we change the sign of the poloidal flow. The two fluid model has an intrinsic poloidal flow present, and its behaviour is at variance from helical flow V-RMHD results. In the V-RMHD case \cite{Mendonca2018}, we have an asymmetry in the effect of the helical flows. The present case is considerably more complicated as  there are both imposed axial and poloidal flows, in addition to the intrinsic poloidal flows. In Fig. \ref{helgr1}, we observe the variation of the mode growth rate with increasing axial flow while keeping poloidal flow constant, for different density gradients at a fixed Pr, the difference between the normalized growth rates trends is significant. The trends show a similarity to the axial flow case, but the big difference is that for $\alpha=0$, the curve nearly coincides with $\alpha=0.5$, for $M_{z}$ negative, showing the strong influence of poloidal flow in modifying the behavior of the mode here. The other curves are parallel to each other, instead of coinciding as in the case of pure axial flow. The frequency curves in Fig. \ref{helfr1} show a similar trend to the corresponding pure axial flow case, \ref{axfr1}, however, the frequency curves have a greater distance between them, showing that the poloidal flow has increased mode frequencies, and contributes to symmetry breaking.

\begin{figure}[!htbp]
\includegraphics[scale=0.32]{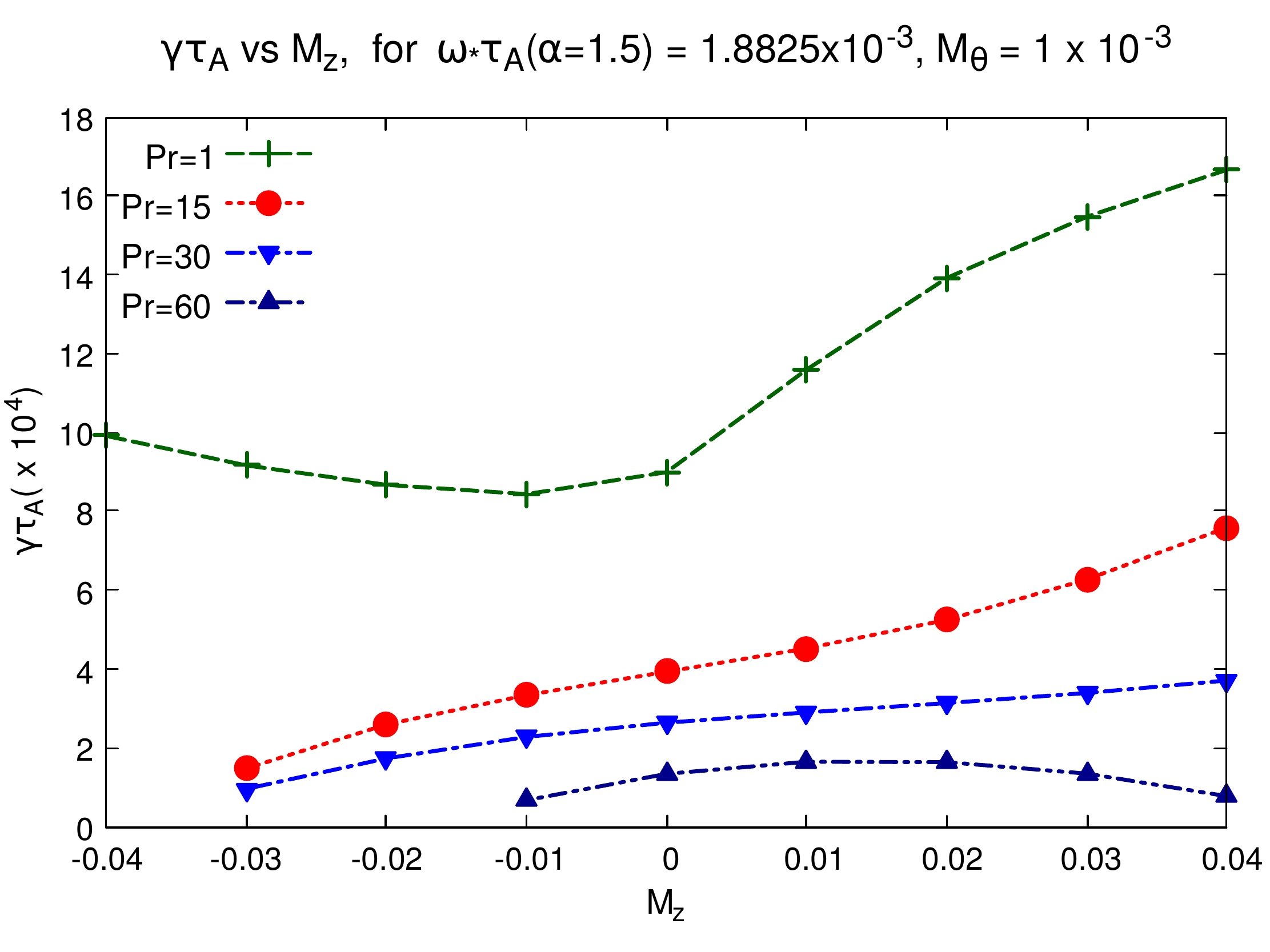} 
    \caption{\small Normalized growth rate vs Helical Flow($M_{z}$,fixed $M_{\theta}\left(\rho_{q=1}\right)=1\times10^{-3}$) for different Pr at a fixed density gradient}
      \label{helgr2}
\end{figure}

In Fig. \ref{helgr2} we present the variation of the normalized growth rate with $Pr$. We observe a similarity with the corresponding axial flow case Fig. \ref{axgr2}, with the addition of poloidal flow only changing the relative distance between the curves. This indicates that the axial flow dominates the dynamics in the case of using combined imposed flows. However, the effect of poloidal flows is also noticeable, despite being minor. 

\begin{figure}[!htbp]
\includegraphics[scale=0.32]{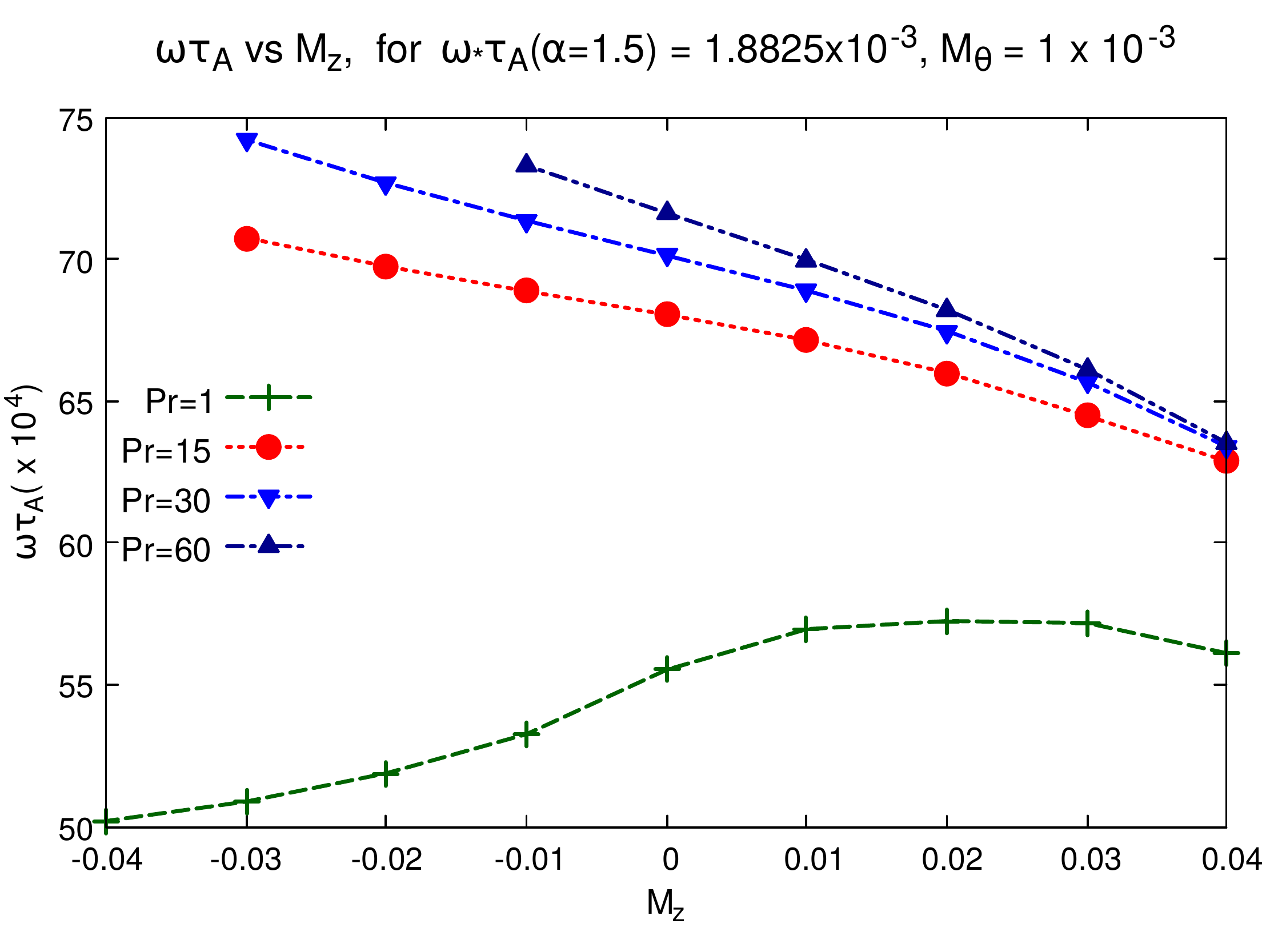}  
     \caption{\small Frequency vs Helical Flow($M_{z}$,fixed $M_{\theta}\left(\rho_{q=1}\right)=1\times10^{-3}$) for different Pr at a fixed density gradient}
      \label{helfr2}
\end{figure}

A similar situation is noticed in the case of Fig. \ref{helfr2}, which is very similar to \ref{axfr2}, showing that the axial flow dominates the dynamics of the system. The frequency curves are very similar, and the addition of imposed poloidal flow seems to only slightly widen the distance between the curves.

In conclusion, we have discussed the effects of imposing an axial and poloidal flow simultaneously on the (1,1) kink mode. We see that overall, axial flow dominates the dynamics, but poloidal flow has a smaller but noticeable effect. The figures are very similar to those obtained with a pure axial flow in the system. 

\clearpage

\section{Nonlinear Studies}\label{nonlinear results}

We present results of our nonlinear simulations in this Section. The nonlinear results here differ from the nonlinear results we obtained in our V-RMHD case\cite{Mendonca2018}. It is worth mentioning here that the nonlinear results differ from the linear in a non trivial sense, which is to say, due to the coupling of the equations, it is not possible to isolate the effect of the individual nonlinear terms. There is an interaction of the nonlinear terms, which cannot be captured by an asymptotic analysis, and a full solution yields results qualitatively different from what an asymptotic analysis would yield. Therefore a numerical solution of the equations is the only way we can understand the true dynamics of this system.  There is a discussion about these issues in Thyagaraja et al\cite{Thyagaraja2000}. Profile-fluctuation interactions as described are the strength of the CUTIE code, and enables us to understand the long term evolution of the visco-resistive modes better. Further, in our system, nonlinear mode coupling is allowed, but we have held the profiles of the equilibrium flows  constant during the evolution. We describe below the variation of the energy levels at saturation as a function of Pr. We observe that at low Pr, the general trend is that the saturated energy level increases with $\alpha$, and also as a function of Pr. At a high Pr, here $Pr>15$, we see that the saturated energy levels fall with increasing $\alpha$.

\begin{center}
\begin{figure}[!htb]
\centering
\includegraphics[scale=0.4]{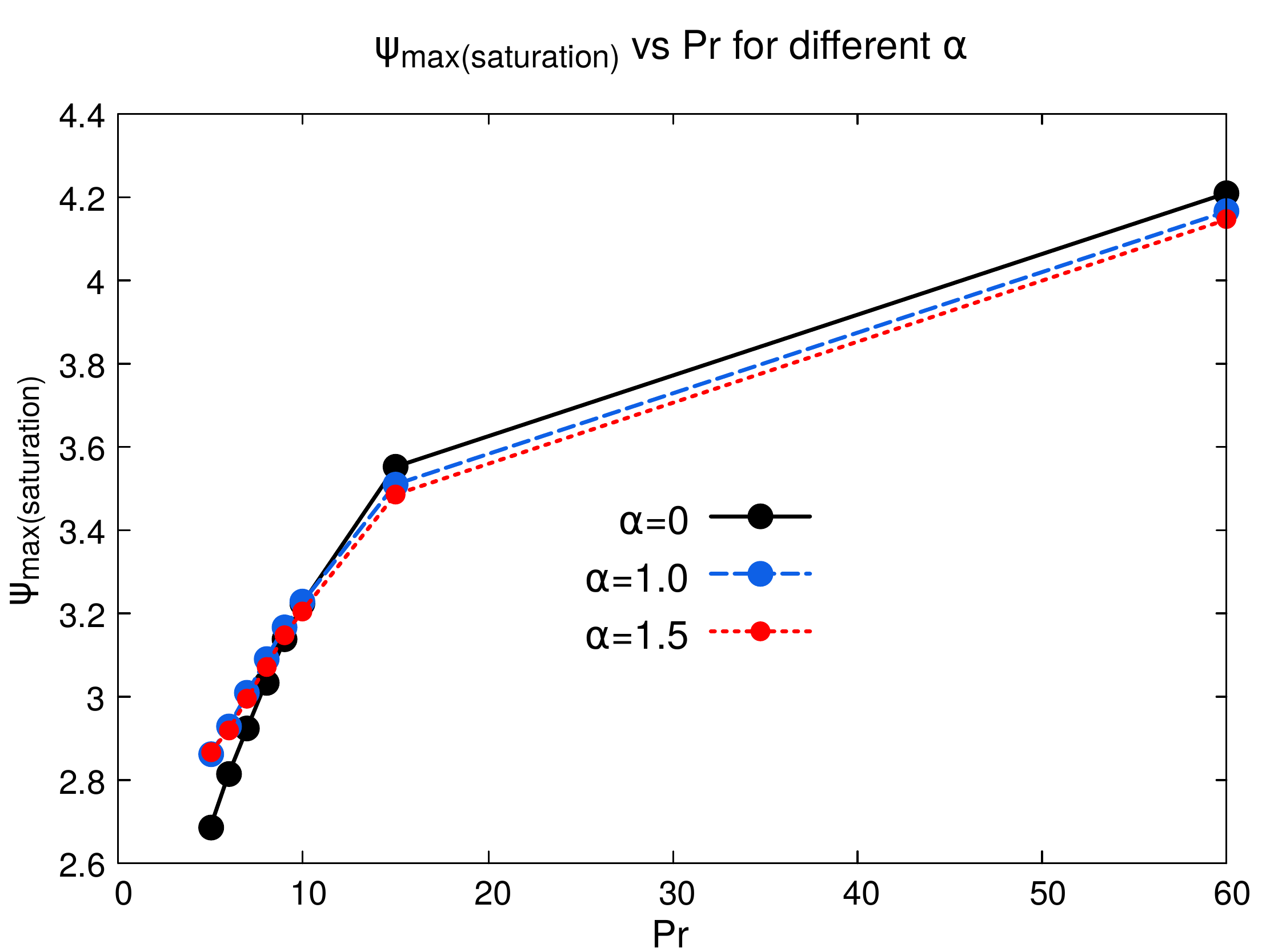} 
\caption{$|\psi_{max}|$ ($|\frac{dB{r}}{(B_{0})^{2}}|$) at saturation vs $Pr$ for different $\alpha$, and thus corresponding $\omega_{*}\tau_{A}(\alpha)=1.255\times\alpha\times1 \times10^{-3}$}
\label{fig:nonlinear saturation study}
\end{figure}
\end{center}

\begin{figure}[!htbp]
\centering
  \begin{minipage}[b]{0.45\textwidth}
     \includegraphics[width=\textwidth]{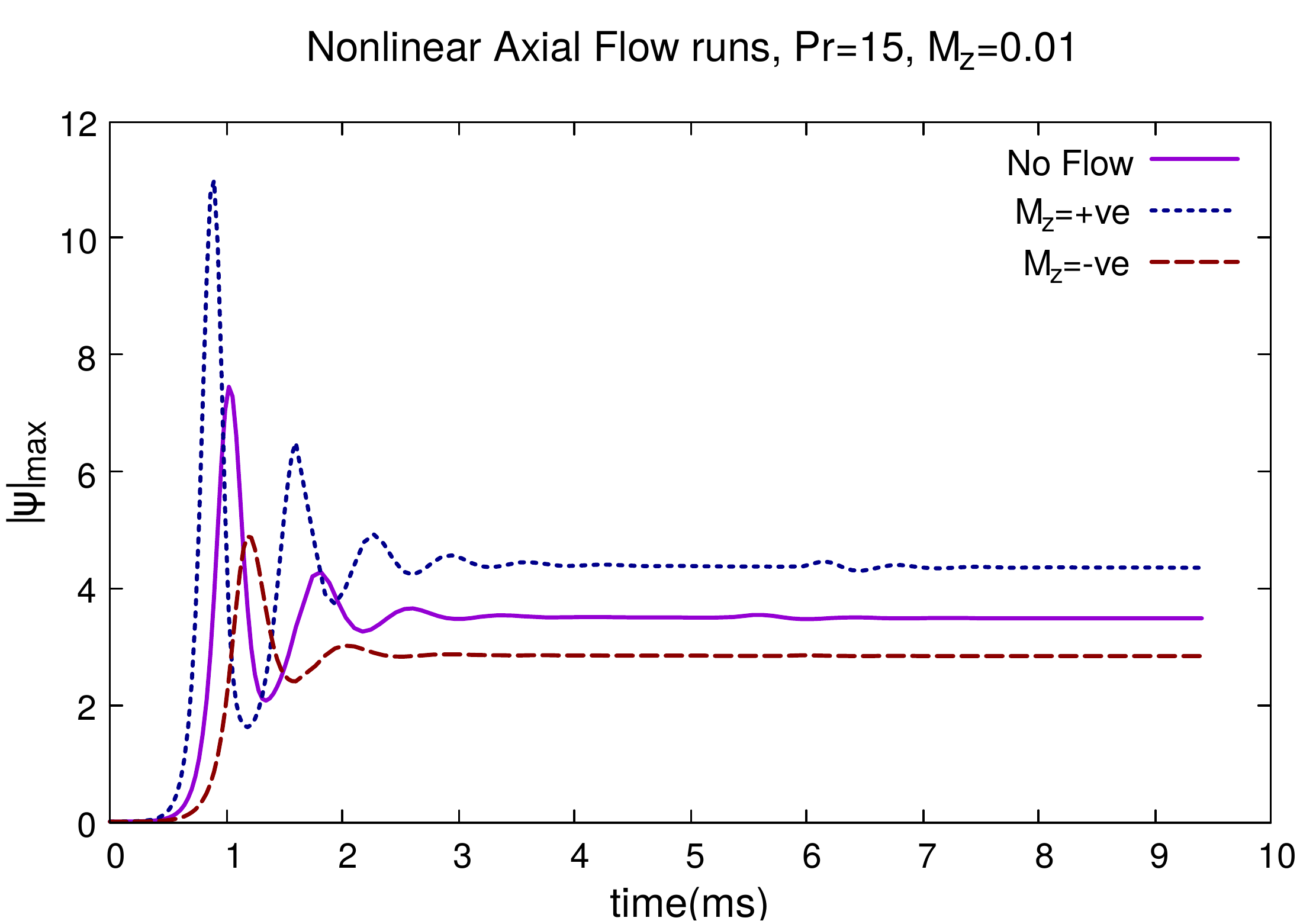} 
     \caption{\small Nonlinear Axial Flow results for Pr=15 } 
            \label{axnl1}
  \end{minipage}
\hfill  
  
   \begin{minipage}[b]{0.45\textwidth}
      \includegraphics[width=\textwidth]{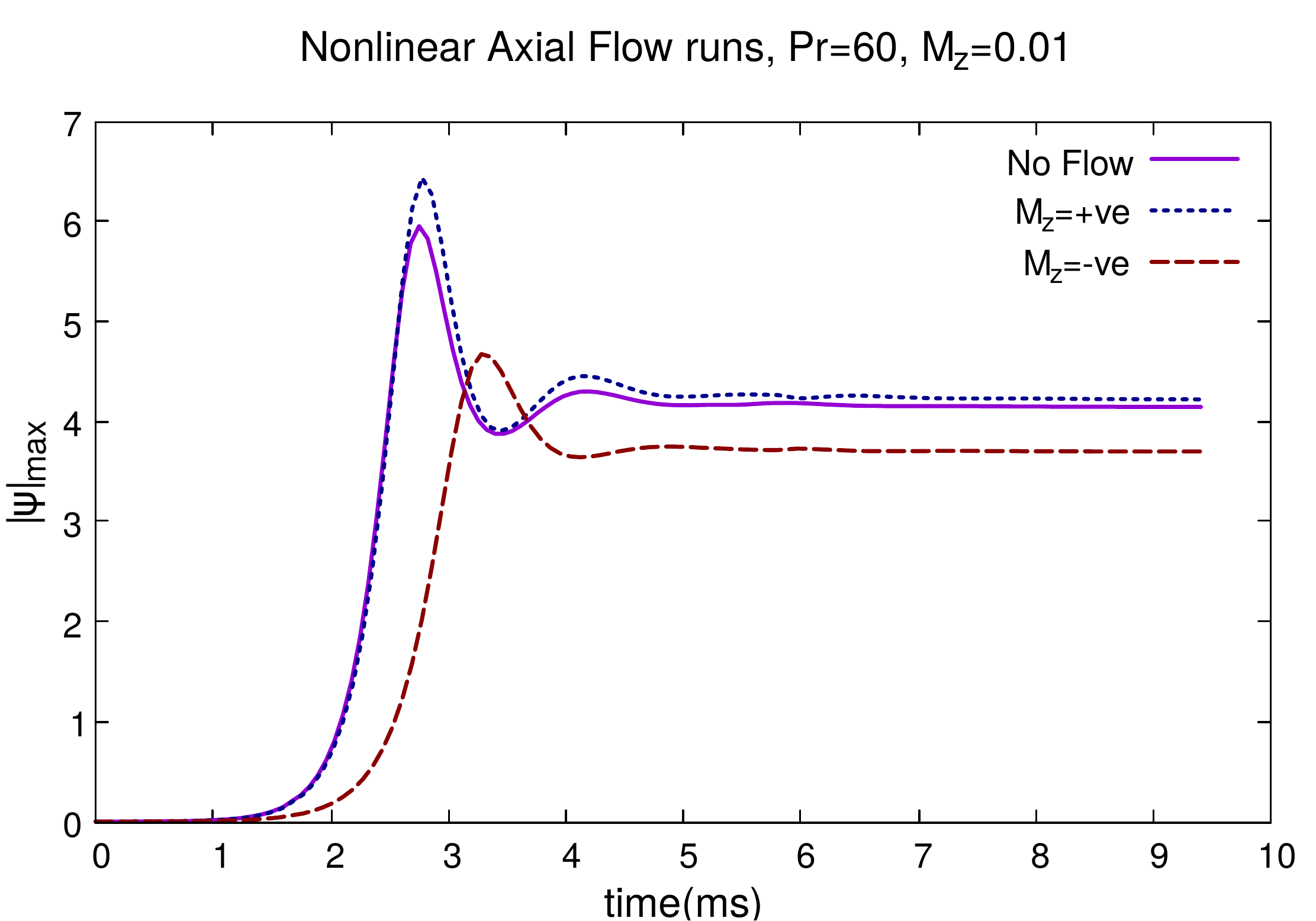} 
         \caption{\small Nonlinear Axial Flow results for Pr=60 }    
           \label{axnl2}
  \end{minipage}

\end{figure}

The nonlinear runs presented in this section were all carried out with a $\alpha=1.5$, thus the $\omega_{*}\tau_{A}(\alpha)=1.8825\times1 \times10^{-3}$ .
In  Fig.  \ref{axnl1}, we observe the nonlinear evolution of $|\psi|$, for the case with imposed axial flow with a fixed Mach number but opposite directions in comparison with the no flow case for $Pr=15$. We notice that positive axial flow leads to a higher initial rise of magnetic fluctuation amplitude as compared to the no flow case, followed by certain oscillations leading to a higher saturation amplitude. The situation is reversed with a negative axial flow, leading to a lower saturation level than the no flow case. 

In the Fig. \ref{axnl2}, we examine the case when we change the Pr to 60. As opposed to $Pr=15$, at the higher viscosity, positive flow hardly differs from the no imposed axial flow case, whereas the saturated amplitude is significantly lower than the no flow case. However, the saturated amplitudes in Fig. \ref{axnl2} have higher values than the corresponding ones in Fig. \ref{axnl1}, consistent with Fig. \ref{fig:nonlinear saturation study}. If we compare these results with our previous results with V-RMHD, we see that in the V-RMHD case, at low viscosity, axial flow cases had higher linear growth rates and saturated amplitudes than the no flow case. However, the high viscosity case, there at $Pr=100$, look similar to the high viscosity case here at $Pr=60$.

\newpage

In the next figure, Fig.  \ref{polnl1} we observe the nonlinear evolution of $|\psi|$, for the case with imposed poloidal flow with a fixed Mach number, $M_{\theta}\left(\rho_{q=1}\right)=1\times10^{-3}$ but opposite directions in comparison with the no flow case for $Pr=15$, we examine the cases with an imposed poloidal flow. Here, the negative imposed poloidal flow case shows destabilisation, that is the saturated amplitude and linear growth rate is higher than the no flow case. It is the opposite for case with imposed positive poloidal  flow, the saturated amplitudes and linear growth rates are lower than the no flow case but to a lesser extent.

For the $Pr=60$ case, in Fig \ref{polnl2} we observe that the positive flow exhibits similar behaviour as in Fig. \ref{polnl1} but the negative flow also shows stabilisation, that is a smaller saturated amplitude than the no flow case, but very slightly, as compared to the larger stabilisation shown in the positive flow case.

\begin{figure}[!htbp]
\centering
  \begin{minipage}[b]{0.45\textwidth}
     \includegraphics[width=\textwidth]{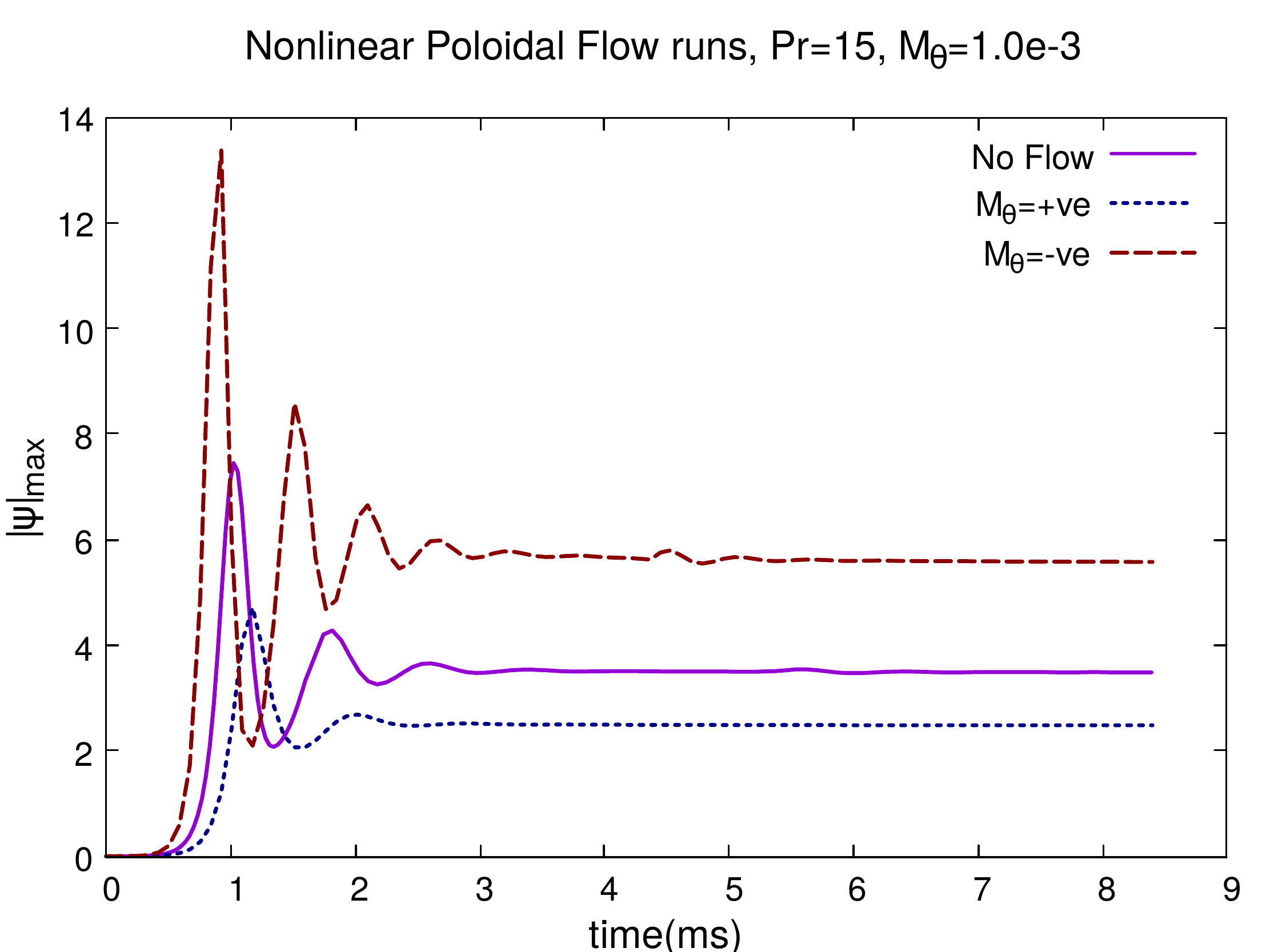} 
     \caption{\small Nonlinear Poloidal Flow results for Pr=15 }  
       \label{polnl1}
  \end{minipage}
  \hfill
  \begin{minipage}[b]{0.45\textwidth}
     \includegraphics[width=\textwidth]{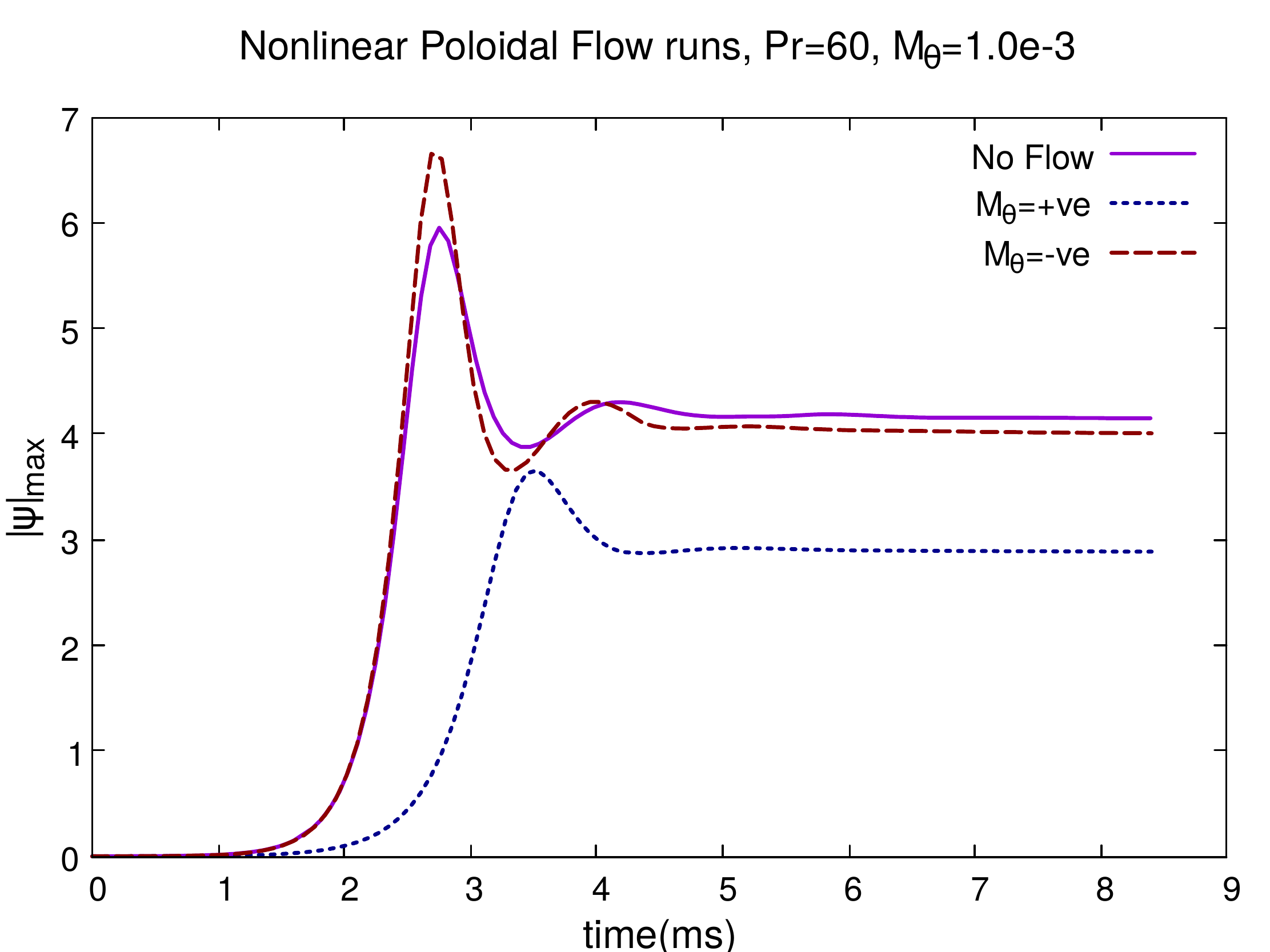} 
       \caption{\small Nonlinear Poloidal Flow results for Pr=60 }  
         \label{polnl2}
  \end{minipage}

\end{figure}

\newpage

\begin{figure}[!htbp]
\centering
  \begin{minipage}[b]{0.45\textwidth}
     \includegraphics[width=\textwidth]{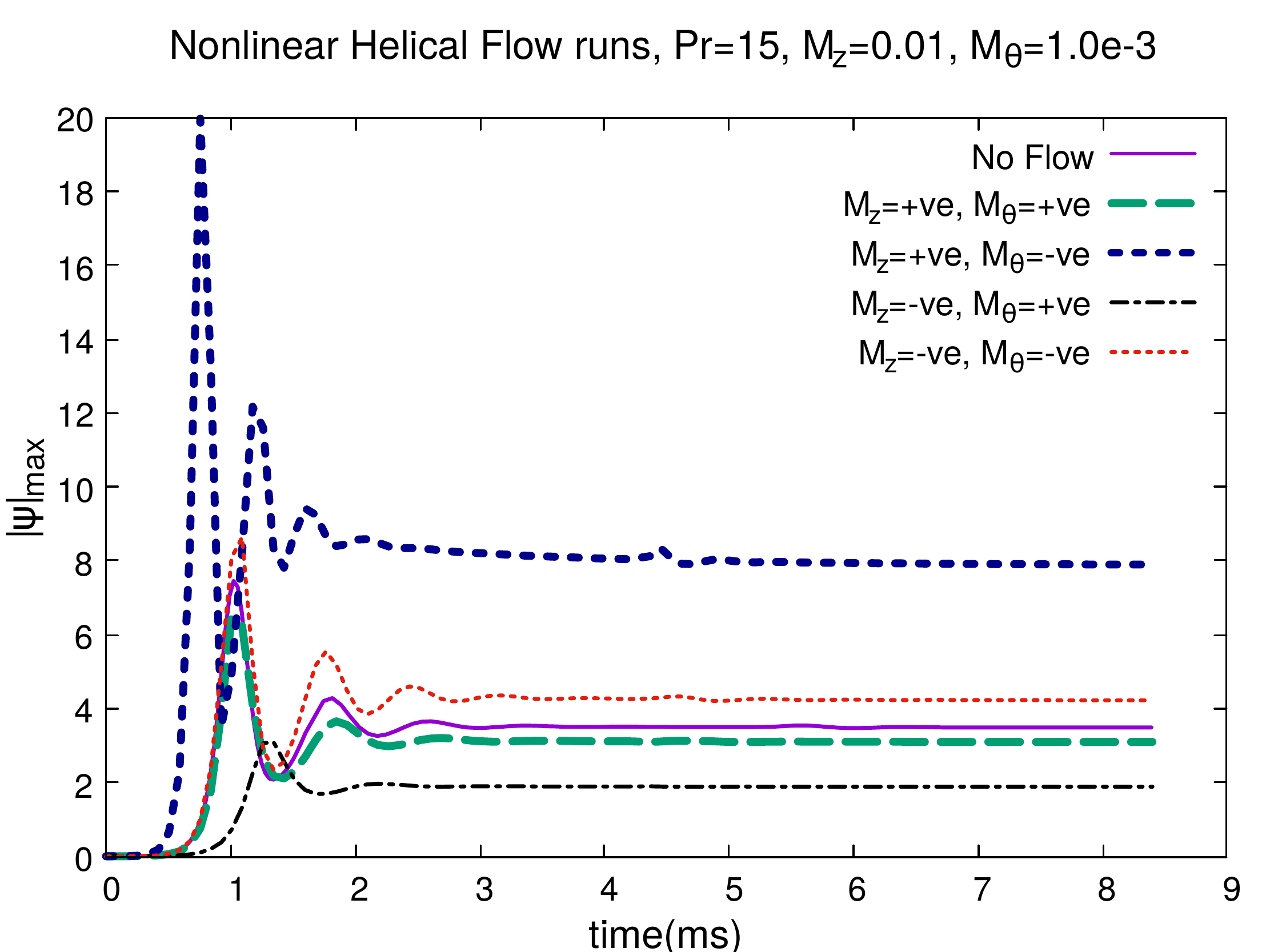}    
      \caption{\small Nonlinear Helical Flow results for Pr=15 }
       \label{helnl1}
  \end{minipage}
  \hfill
  \begin{minipage}[b]{0.45\textwidth}
    \includegraphics[width=\textwidth]{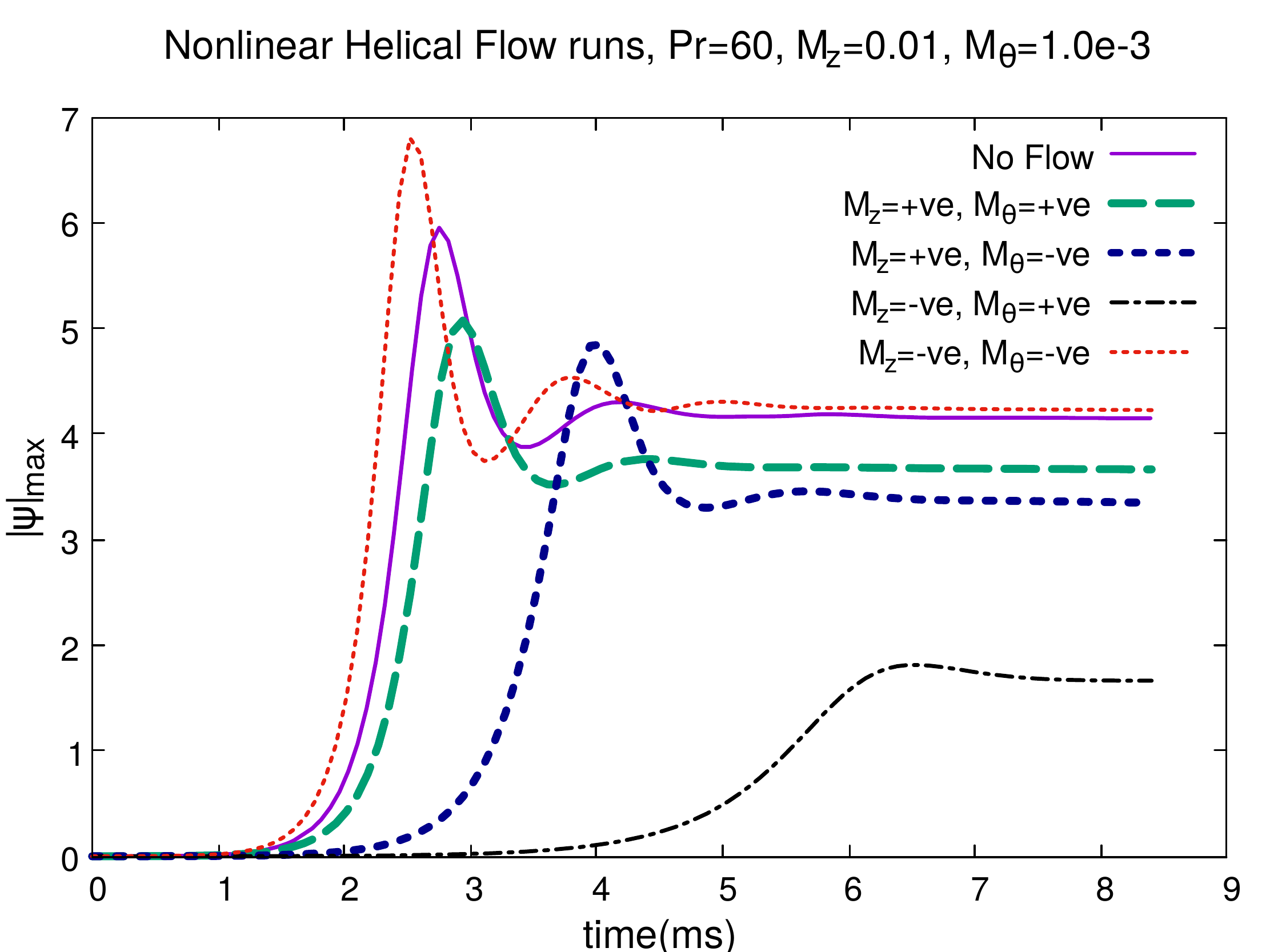} 
      \caption{\small Nonlinear Helical Flow results for Pr=60 }  
        \label{helnl2}
  \end{minipage}

\end{figure}

Finally, we turn to the cases with combined imposed flows. In Fig. \ref{helnl1}, where we observe the nonlinear evolution of $|\psi|$, with helical flows, that is, we use the positive and negative combinations of a fixed axial Mach number $M_{z}=0.01$ and poloidal Mach number $M_{\theta}\left(\rho_{q=1}\right)=1\times10^{-3}$. This results in four cases, as there are two cases each of axial and poloidal flow. An examination of the Fig. \ref{helnl1} shows us that it exhibits features of both the nonlinear axial flow Fig. \ref{axnl1} and Fig. \ref{polnl1} for the same $Pr$. We notice an asymmetry having runs with two helicities destabilised, that is, with higher saturated amplitudes, and two cases more stable than the no flow case. As in Fig. \ref{polnl1}, the cases with $M_{\theta}\left(\rho_{q=1}\right)$ positive are stabilized and vice versa. Negative axial flow is stabilizing and positive axial flow destabilizes if we fix the poloidal flow. This is similar qualitatively to what we had seen in Fig. \ref{axnl1}. This also indicates that the intrinsic poloidal flow and imposed poloidal flow have the same effects on the mode and it is the overall poloidal flow, the sum of both which affects the linear and nonlinear characteristics of the mode. In the V-RMHD, we had obtained a destabilization in all cases of imposed helical flow, unlike the present case which is more complicated.

The behavior of the mode in the presence of helical flow in Fig. \ref{helnl2} similarly exhibits the features of Fig. \ref{axnl2} and Fig. \ref{polnl2}. Here, for positive poloidal flow, positive axial flow is destabilizing but for negative poloidal flow, negative axial flow is destabilizing. Conversely, for a fixed negative axial flow, positive poloidal flow is stabilizing while for a fixed positive axial flow, positive poloidal flow is destabilizing. In comparison to the $Pr=15$ case, we see that case with positive axial and negative poloidal flow which had the highest saturated amplitude, now comes at third place, while the others remain at the same relative position. In comparison to V-RMHD, we see a similarity in that the most stable case is the one with axial and poloidal flows negative, but the important difference is that in V-RMHD, all the cases had a lower saturated amplitude than the no flow case.  

\newpage

\section{Summary and Discussion}\label{discussion}

In summary, we have carried out linear and nonlinear studies of the $m=1$,$n=1$ mode in a two fluid regime using the CUTIE code. In the absence of any imposed flow, we have observed significant diamagnetic stabilisation in the growth rate of the $m=1$,$n=1$ mode but the effect diminishes as we increase the viscosity. In low viscosity, the mode rotation frequencies are smaller compared to the respective diamagnetic flow frequencies but they approach the diamagnetic flow frequency asymptotically as we increase the viscosity. This is a novel result highlighting a reactive contribution of viscosity that influences the real frequency of the resistive kink mode. We have also found that the numerical results can be accurately fitted to a scaling relation given by  $\omega\tau_{A}\propto\frac{(Pr-0.5)}{Pr+1.5}$, where the proportionality constant changes with the $\alpha$ value.  This scaling relation is found to provide a good analytic fit to the data points for all values of $\alpha$. Our findings can form the basis of a future theoretical investigation to obtain a first principles model of this behavior.
 The nonlinear saturation levels of the modes also increase with higher viscosity values, however its behaviour with change in diamagnetic flow is different for the low viscosity and high viscosity regime. At low viscosity it increases slightly with diamagnetic flow but decreases for high viscosity. In case of imposed axial flow, there is asymmetry with respect to the direction of the axial flow, for all cases, including zero diamagnetic flow. This is because parallel dynamics also introduce an asymmetry in the linear growth rate other than diamagnetic flow. Unlike axial flow, in the case of poloidal flow, there's a symmetry in the growth rate for the zero diamagnetic flow case. However, after the introduction of diamagnetic flow, the growth rates are not symmetric with respect to changing the direction of the imposed poloidal flow. In one direction it is destabilising and in the other direction it is stabilising the mode. This is very different from the single fluid result that the poloidal flow is always stabilising and recent experimental observations in JET\cite{crombe} indicate that the poloidal flow may destablise the internal kink mode for certain conditions. Further extensions of these results in a fully toroidal code are presently in progress and will be reported in future.\\

\section{Acknowledgement}
AS is thankful to the Indian National Science Academy
(INSA) for their support under the INSA Senior Scientist Fellowship
scheme.
\section{References}
\bibliographystyle{ieeetr}
\bibliography{paper_2fluid_nf_draft_r}

\end{document}